\documentclass[fleqn,usenatbib]{mnras}

\usepackage{newtxmath,newtxtext}

\usepackage[T1]{fontenc}

\DeclareRobustCommand{\VAN}[3]{#2}
\let\VANthebibliography\thebibliography
\def\thebibliography{\DeclareRobustCommand{\VAN}[3]{##3}\VANthebibliography}

\usepackage{graphicx}	
\usepackage{amsmath}	
\usepackage{amssymb}	

\usepackage[FIGTOPCAP,Large]{subfigure}
\usepackage{cancel}

\title[Are there pristine comets?]{Are there any pristine comets? Constraints from pebble structure}

\author[Malamud et Al.]{
	Uri Malamud$^{1,2}$\thanks{E-mail: urimala@physics.technion.ac.il},
	Wolf A. Landeck$^{3}$,
	Dorothea Bischoff$^{3}$,
	Christopher Kreuzig$^{3}$,
	Hagai B. Perets,$^{1}$,
	\newauthor
	Bastian Gundlach$^{3}$
	and J\"urgen Blum$^{3}$
	\\
	${^1}$Department of Physics, Technion - Israel Institute of Technology, Technion City, 3200003 Haifa, Israel\\
	${^2}$School of the Environment and Earth Sciences, Tel Aviv University, Ramat Aviv, 6997801 Tel Aviv, Israel\\
	${^3}$Institute for Geophysics and extraterrestrial Physics, Technische Universität Braunschweig, Mendelssohnstr. 3, D-38106, Braunschweig, Germany\\
}

\pubyear{2022}

\begin{document}
	
    \label{firstpage}
	\pagerange{\pageref{firstpage}--\pageref{lastpage}}
	\maketitle
	
	\begin{abstract}
	We show that if comets (or any small icy planetesimals such as Kuiper belt objects) are composed of pebble piles, their internal radiogenic as well as geochemical heating results in considerably different evolutionary outcomes compared to similar past studies. We utilize a 1D thermo-physical evolution code, modified to include state-of-the-art empirical measurements of pebble thermal conductivity and compression, the latter obtained through a new laboratory experiment presented here for the first time. Results indicate that due to the low pebble thermal conductivity, the peak temperatures attained during evolution are much higher than in any previous study given the same formation time. Assuming meteoritic radiogenic abundances, we find that only extremely small, sub-kilometre comets have the potential to retain the primordial, uniform and thermally unprocessed composition from which they formed. Comets with radii in excess of about 20 km are typically swept by rapid and energetically powerful aqueous hydration reactions. Across the full range of comet sizes and formation times, evolutions result in the processing and differentiation of various volatile species, and a radially heterogeneous nucleus stucture. Our computations however also indicate that the assumed fraction of radionuclides is a pivotal free parameter, because isotopic analyses of the only available cometary samples suggest that no $^{26}$Al was ever present in comet 81P/Wild 2. We show that if comets formed early in the protoplanetary disc (within 1-3 Myr), the radionuclide abundances indeed must be much smaller than those typically assumed based on meteoritic samples. We discuss the importance of our findings for the formation, present-day attributes and future research of comets.\\
	\end{abstract}
	
    
	\section{Introduction}\label{S:Intro}
	In recent years, there has been an increasing interest in comet formation scenarios advocating the gravitational collapse of pebble clouds \citep{Youdin.2005,JohansenEtAl-2007,BlumEtAl-2014}. Pebbles are mm to cm aggregates composed of smaller, micrometer-sized dust or ice constituent particles \citep{ZsomEtAl-2010,LorekEtAl-2018}. The structure of comets formed by pebble-cloud collapse is hierarchic. The pebbles are randomly packed one on top of the other, with voids in between. They form a loosely-bound gravitational pile. The macro-porosity of such a pile, for semi-spherical and similarly-sized pebbles is approximately 40\%. This result is often obtained in laboratory experiments, and it does not vary greatly unless the shape/texture of the pile constituents significantly deviates from spherical symmetry \citep{YuZou-1998}. The pebbles themselves are likewise porous aggregates. They form via extremely low-velocity collisions and hence sticking of their constituent micron-sized grains. Laboratory experiments show that in such conditions the typical pebble intra-porosity is approximately in the range 60-70\% \citep{WeidlingEtAl-2012}.
	
	The effective thermal conductivity of such a hierarchic network of packed pebbles is extremely low, up to 3 or even 4 orders of magnitudes less than that of the constituent pebble grains. This arises from highly reduced contacts between solid elements, first as a result of the pebble intra-porosity, and second due to the pebble packing \citep{GundlachEtAl-2020}. In addition, this hierarchic structure affects the heat transport inside the comet nucleus since energy may be radiated effectively through the void space between the pebbles. While the porosity correction greatly reduces the thermal conductivity, as mentioned above, radiative heat transport has the opposite effect, however it may only increase thermal conductivity by up to a factor of a few \citep{HuEtAl-2019}. Combined, these two effects are known to regulate the near-surface temperature and therefore the activity of comets. In recent years, various studies have indeed looked at different aspects related to the surface of comets, that are associated with these thermal modifications \citep{HuEtAl-2019,GundlachEtAl-2020,Davidsson-2021}. However, much less is known about the role of pebble structure on the internal, long-term physical and thermal evolution of comets.
	
	In this paper we will show that the internal evolution of comets formed by the gentle collapse of a pebble cloud may critically constrain their size, formation time and internal structure. If indeed comets are comprised of pebbles, we can use their internal evolution in order to indirectly infer several important clues about the early formation of the Solar system and the present-day inner distribution of comet composition. Throughout the paper, we use the term 'comet' when referring to any similar-sized, small icy planetesimal. That is, our conclusions are interchangeable for all similar-sized icy planetesimals, including e.g. Kuiper belt or Oort cloud objects, but the word 'comet' is hereafter used exclusively, for simplicity.
	
	The internal evolution of comets has been extensively investigated over the last several decades. While the thermal and in turn physical modifications inside comets are governed by several competing timescales \citep{HuebnerEtAl-2006}, the most important of those are the radiogenic heating timescale and the thermal timescale. The former dictates how quickly heat is released by the decay of radioactive species, whereas the latter dictates how quickly heat can escape over a particular length scale.
	
	Early studies by \cite{WhippleStefanik-1965} and \cite{PrialnikEtAl-1987} have reached the conclusion that long-lived radioactive (LLR) species such as $^{40}$K, $^{232}$Th, $^{235}$U and $^{238}$U are capable of at most heating the comet to a peak temperature of a few dozen Kelvin. On the other hand, short-lived radioactive (SLR) species such as $^{26}$Al and $^{60}$Fe are considered a more powerful heat source if the comet formation time is sufficiently short relative to the formation of Calcium Aluminium Inclusions (CAI), which increases their initial abundance \citep{PrialnikEtAl-1987}. In particular, \cite{PrialnikPodolak-1995} have shown that the peak temperature during the early evolution of comets can attain very large values, even beyond the melting temperature of water, but the thermal conductivity must be sufficiently low, otherwise heat would be transported to the surface quickly and the temperature could not build up. Also, the formation time of the comet has to be adequately short, increasing the abundance of SLRs, and thus supplying enough energy to significantly raise the temperatures. Hot interiors typically required the comets to be very sizeable, with radii over tens or even hundreds of km \citep{HaruyamaEtAl-1993,Yabushita-1993}, since the thermal timescale depends not only on the thermal conductivity but also primarily on the size. 
	
	Given the state of knowledge at the time, it was difficult to justify extremely low thermal conductivities. The study by \cite{PrialnikPodolak-1995} nevertheless explored toy models with a thermal conductivity that was 3 orders of magnitudes lower than the typical values considered appropriate at the time, even lower than those of amorphous ice. It was shown that for comets with radii over 20 km, the peak temperature was relatively high ($>$100 K) and essentially unaffected when lowering thermal conductivity (the thermal timescale in their model was sufficiently large either way), however the cooling time of the comet became much larger, retaining high internal temperatures for at least hundreds of Myr.
    
    More recent studies \citep{ChoiEtAl-2002,MerkEtAl-2002,MerkPrialnik-2003,MerkPrialnik-2006,DeSanctisEtAl-2007,MousisEtAl-2012,HolmEtAl-2015,MousisEtAl-2017,GolabekJutzi-2021} have built on earlier work, adding various details and relations. In the last couple of decades most studies have newly included SLR species $^{60}$Fe in addition to $^{26}$Al, however with the exception of a few \citep{ChoiEtAl-2002,DeSanctisEtAl-2007} they have also typically neglected LLRs, based on the conclusions of early studies that they are ineffective heat sources in comets. All the evolution studies dedicated to comets, both early and recent, generally predict outcomes in which liquid water is not present, since sufficiently low thermal conductivities have not yet been robustly and self-consistently invoked (i.e motivated theoretically and empirically), and since comet radii rarely reach beyond several kilometres. Alternatively, two studies by \cite{WickramasingheEtAl-2009} and \cite{HolmEtAl-2015} that focus on the astrobiological implications of liquid water in comets, argue that it is possible to obtain liquid water, however the formation time they invoke is extremely small and on the order of 1 Myr.
    
    We also note for completion that the literature includes many studies that have targeted larger water-bearing planetesimals, in the size range that covers satellites and dwarf planets (which are certainly large enough to retain more heat). These models, brought forward by \cite{SchubertEtAl-2007,PrialnikMerk-2008,DeschEtAl-2009,NeumannEtAl-2015}, are designed to compute the process of internal differentiation, however each utilising a considerably different conceptual approach. Each of these models have nowadays given rise to dozens of new off-spring studies, incrementally introducing various improvements and broadening their scientific objectives.
    
    As a recent example, consider the study of \cite{CarryEtAl-2021} aiming specifically at main belt objects that are potentially water-bearing. They show that having a radius of 135 km, main-belt asteroid (87) Sylvia, is probably a differentiated, water-bearing body. They however conclude that if they apply their model to (considerably smaller) comets, the outcome is an unaltered, pristine composition. The underlying reason is that the effective thermal conductivity is insufficiently low, given their assumed porous structure (see their figure 6). Other models, such as that of \cite{LichtenbergEtAl-2021} and references therein, focus on the heating by $^{26}$Al of extremely early-formed planetesimals ($<3$~Myr after the formation of CAI), their compositional fractionation, and consequent contribution to planet formation. This study invokes pebble accretion, but the thermal conductivity used for unsintered porous silicates is parametrised to a single value, rather than accounting for the complex relation that depends on temperature, structural and other relevant parameters. This approach might be justified given the study's objectives, but is ill-suited for detailed comet evolution modeling.
    
    Given the previous overview of the existing literature, our study is unique in several aspects: (a) We incorporate both SLRs and LLRs, and show that LLRs may not be ignored in large comets formed by pebble-cloud collapse; (b) Our treatment of the solid state thermal conductivity is based on experiments that are specifically tailored to porous pebble media \citep{GundlachBlum-2012}, whereas other studies have used various other constituent relations. In addition, we also newly consider a radiative transfer term for the net thermal conductivity, since many modern codes \citep{HuEtAl-2019,GundlachEtAl-2020,Davidsson-2021} have shown radiative transfer to be of comparable contribution and even dominate over solid state conductivity, depending on temperature and pebble size; (c) We use a fully self-consistent thermo-physical model which considers the phase transitions during exothermic hydration (serpentinization) reactions that occur when liquid water and rock interact chemically. Such interactions have very significant implications since they provide a powerful additional heat source, and change the rock structural and thermal properties. However, they are typically associated with larger icy objects such as moons and dwarf planets \citep{MalamudPrialnik-2013,MalamudPrialnik-2015,NeveuDesch-2015,MalamudPrialnik-2016,DeschNeveu-2017,MalamudEtAl-2017,MalamudPerets-2018} and have so far been ignored in comet modeling. Here we show that even relatively small, pebble-pile comets, may be sufficiently heated to obtain liquid water and trigger aqueous alterations throughout most of the interior. We postulate that the presence of liquid water in turn triggers the instantaneous collapse of the hierarchic pebble structure, which we follow in detail; (d) We also perform new laboratory measurements and determine the reduction of the pebble packing porosity due to self-gravity. We use these empirical fits to calculate accurate initial density profiles; (e) We newly consider reduced radiogenic heating, if radionuclides are only incorporated into refractory minerals, whose fraction (among the 'rocky material') could be smaller than 1. According to \cite{BardynEtAl-2017}, approximately half of the refractories in comet 67P/C–G could be in organics, greatly reducing the amount of available radiogenic heat. Much lower fractions still are implied by performed isotopic analyses of Stardust samples \citep{Levasseur-RegourdEtAl-2018}, consistent with the meteoritic mineral fraction inferred by Rosetta \citep{FulleEtAl-2017}; and (f) We consider various permeability coefficients for the flow model, taking after the two limiting cases suggested by \cite{GundlachEtAl-2020}, since the exact permeability for pebble structure is still an open topic of research.
    
    The paper layout proceeds as follows. Model details and caveats are presented in Section \ref{S:Model}; In section \ref{S:ParameterStudy} we perform a large parameter study, depicting the thermal consequences of evolution as a function of comet size, formation time and various other key parameters; We follow with a detailed analysis of various emerging archetypal evolutionary classes in Section \ref{S:Archetypes}; The results are then discussed in Section  \ref{S:Discussion} and summarised in Section \ref{S:Summary}.
    
    \section{Model details and caveats}\label{S:Model}
    For this paper we use the code \emph{SEMIO} (structure and evolution modeling of icy objects). The code was presented in full by \cite{MalamudPrialnik-2016}, and we refer to this manuscript and the references therein for a complete review of its capabilities and design, equations, parameters and numerical scheme. Several new features are added, including a more accurate, temporally evolving host-star luminosity, for the minor planet boundary condition \citep{MalamudEtAl-2017}. Other upgrades, modifications and constituent relations are applicable specifically to comets formed by pebble-cloud collapse and will be fully presented below in Section \ref{SS:Updates}.
    
    \subsection{Model overview}\label{SS:Overview}
    We start with a short summary of the code's capabilities and attributes. Our evolution model \emph{SEMIO} is designed to couple the thermal, physical and chemical evolution of icy minor planets, in a fully self-consistent manner, and uses a dynamic time step for long term (Gyr) evolutionary computations. It considers the energy contribution primarily from radiogenic heating, latent heat released by crystallisation of initially amorphous ice, energy released by aqueous geochemical serpentinization (hydration) reactions and surface insolation. It treats heat transport by conduction and advection, and follows the transitions among four phases of water (amorphous ice, crystalline ice, liquid and vapour), and two phases of silicates (anhydrous rock and hydrous rock). The model considers the contribution/absorption of energy by phase transitions among the various phases of water (condensation / deposition / freezing / sublimation / evaporation / melting), noting that when these contributions enter the total energy budget they are actually of minor significance compared to other internal heat sources. When the internal temperature becomes sufficiently high (typically above 670-700 K), the rock may undergo the reverse process of serpentinization (also known as dehydration), in which the rock exudes the water it had previously absorbed. Despite exploring a very large parameters space, we note that temperatures are never that high in our current study.
    
    Our flux equations are solved for water gas and liquid, based on the multiphase flow mechanism first introduced by \cite{PrialnikMerk-2008}. The rocky component is stationary and cannot migrate throughout the object, nor accompany any of the aforementioned water fluxes. The only exception to the previous statement is for the outer boundary, where the code is capable of handling activity-driven pebble ejections at or slightly below the surface. Such activity is however irrelevant as we deliberately place all comets far from the Sun at 40 AU, arbitrarily giving them a Kuiper belt origin with a corresponding initial temperature of $\sim$40 K. We also assume a circular orbit, since eccentric orbits result in small time steps and hence cannot resolve the long-term evolution. 
    
    Finally, the code uses an adaptive-grid numerical scheme, allowing for a change in body size and mass without loss of accuracy. Although our objects are placed far from the Sun and never become active, hence total mass is conserved, they can still change in volume due to compression of their internal pore space. Compression may occur in two ways. First, when self-gravity reduces the packing porosity of the pebbles. In Section \ref{SSS:PebbleCompression} we present new experimental results for pebble packing compression, and based on these data we calculate an accurate initial density profile for each model. For the comet size range considered here, we show that the resulting difference in radii due to self-gravity compaction is small, and only relevant for radii exceeding $\sim$10 km. Second, we assume that full and instantaneous aqueous collapse of pebbles occurs if liquid water appears locally anywhere inside the comet. In such cases the packing porosity of the pebbles completely vanishes, and we are left with a non-hierarchial, homogenous aggregate of the constituent micron-sized grains, wherein the porosity is much lower and the typical pore size is greatly reduced.
    
    As described in Section \ref{S:ParameterStudy}, we perform a parameter study that investigates various model realisations based on a combination of comet sizes, formation times, pebble sizes, mineral fractions and permeability coefficients. The number of possible combinations is extensive and in order to present comprehensible results we otherwise choose singly-defined characteristic values and relations to all other model parameters. Our constituent relations are generally based on values derived for comet 67P/C–G as it is by far the most well-studied cometary archetype \citep{FulleEtAl-2016,FilacchioneEtAl-2019,GroussinEtAl-2019}. These relations include an initial rock/ice mass ratio of 4, albedo of 0.062 and initial uniform bulk density of 533 kg $\times$ m$^{-3}$. We adopt a uniform initial temperature of $T_0=40$ K, which is around the equilibrium surface temperature expected at 40 AU.
    
    \subsection{Model updates}\label{SS:Updates}
    The new model updates address the consequences of pebble structure, as well as refine the most important constituent relations.
    \subsubsection{Thermal conductivity}\label{SSS:ThermalConductivity}
    According to \cite{GundlachEtAl-2020}, the effective thermal conductivity $K_{\rm eff}$ in a structure made of pebbles has two components. The first component is the net solid state conduction term $K_{\rm net}$ which is affected by the hierarchical structure of the pebbles, according to the network of contacts between solid particles, both inside the pebbles (i.e among its constituent grains) and between the pebbles. The second component is the radiative transport term $K_{\rm rad}$, which depends on the mean free path of the photons. The latter in turn depends on various parameters, but most sensitively on the pore size \citep{Merrill-1968}. Hence, only when the pebbles are large and the voids between the pebbles are large - the radiative transport term becomes more significant.
    
    The network thermal conductivity is given by \cite{GundlachBlum-2012}. It can be written as the product of the thermal conductivity of the pebbles $K_{\rm peb}$ and the Hertz factor of the pebbles $H_{\rm peb}$:
    \begin{equation}\label{EQ:net_thermal_conductivity}
    	K_{\rm net} \, = \,  K_{\rm peb}  \, H_{\rm peb}                               \ \mathrm{.}
    \end{equation}

	The thermal conductivity of the pebbles $K_{\rm peb}$ is in itself a product of the thermal conductivity of the intra-pebble grains $K_{\rm gra}$ and the intra-pebble Hertz factor $H_{\rm gra}$, calculated for the constituent grains:
	
	\begin{equation}\label{EQ:pebble_thermal_conductivity}
		K_{\rm peb} \, = \,  K_{\rm gra}  \, H_{\rm gra}                               \ \mathrm{.}
	\end{equation}

    The Hertz factor is denoted by $H_j$, where the subscript $j$ is replaced by either $_{\rm peb}$ or $_{\rm gra}$, respectively, and has the same form for both hierarchies. It is given by
    \begin{equation}\label{EQ:hertz_factor}
    	H_j \, = \,  \left[\,\frac{9\, \left( \, 1 \, - \, \mu_j^2 \, \right)}{4 \, E_j}  \, \pi \, \gamma_j \, r_j^2 \,\right]^{1/3}\frac{f_{\rm 1}  \, \cdot \, \mathrm{exp}(f_{\rm 2}\, \phi_j)}{r_j}   \ \mathrm{,}
    \end{equation}

    \noindent where $\mu$ is the Poisson ratio, $E$ is the Young's modulus, $\gamma$ is the surface energy, $r$ is the particle radius, $\phi$ is the volume filling factor and $f_{\rm 1,2}$ are empirically determined parameters that take the packing geometry into account. For these parameters we follow the prescriptions and values given by Table A1 of \cite{GundlachEtAl-2020}. The only exception is the pebble size $r_{\rm peb}$ which is a free parameter and specified in Section \ref{SS:FreeParameters}.
    
    In order to complete the solid state conduction scheme, we require the bulk thermal conductivity of the pebble grains $K_{\rm gra}$. However, since pebbles may be composed of a (temporally evolving) mixture of various species of rock and ice, we must take the mass weighted average of all different species in order to obtain the bulk
    
    \begin{equation}\label{EQ:Kgrain}
    	K_{\rm gra}(T) = \frac{K_{\rm a}(T)\rho_{\rm a} + K_{\rm c}(T)\rho_{\rm c} + K_{\rm \ell}\rho_{\rm \ell} + K_{\rm u}\rho_{\rm u} + K_{\rm p}\rho_{\rm p}}{\rho_{\rm tot}},
    \end{equation}
    \noindent where amorphous ice is denoted by the subscript (a), crystalline ice (c), liquid water ($\rm \ell$), anhydrous rock (u), hydrous rock (p), $\rho$ denotes the respective densities and $T$ is the temperature. We note that the density $\rho$ is defined \emph{locally} for each component, and should not be confused with the specific density $\varrho$, which is the mass per unit volume of the solid material (i.e. zero-porosity grain density) of each species. For the latter we adopt: $\varrho_{\rm a,c}=917$ kg $\times$ m$^{-3}$, $\varrho_{\rm \ell}=997$ kg $\times$ m$^{-3}$, $\varrho_{\rm u}=3500$ kg $\times$ m$^{-3}$ and $\varrho_{\rm p}=2900$ kg $\times$ m$^{-3}$.
    
    The thermal conductivity for the various species is detailed as follows. For $K_{\rm c}$ we have from \cite{Klinger-1975}:    
    \begin{equation}\label{EQ:Kcrystalline}
    	K_{\rm c}(T)= 567/T~{\rm J} \times {\rm m}^{-1} \times {\rm s}^{-1} \times {\rm K}^{-1}.
    \end{equation}
    
    For $K_{\rm a}$ we have from \cite{Klinger-1980}:
    \begin{equation}\label{EQ:Kamorphous}
    	K_{\rm a}(T)= 2.348\times10^{-3}T+2.82\times10^{-2}~{\rm J} \times {\rm m}^{-1} \times {\rm s}^{-1} \times {\rm K}^{-1}.
    \end{equation}
    
    For the other species $K_{\rm \ell}$, $K_{\rm u}$ and $K_{\rm p}$ we adopt constants. Although they are all temperature-dependent in principle, the variation as a function of temperature is not significant and we prefer constant values for simplicity. For liquid water we adopt the thermal conductivity at the melting temperature, $K_{\rm \ell}=0.55~{\rm J} \times {\rm m}^{-1} \times {\rm s}^{-1} \times {\rm K}^{-1}$, so it may be seen as a lower limit in the temperature range relevant to this study \citep{DincerZamfirescu-2016}. For rocky material, the grain thermal conductivity is highly variable depending on the precise composition assumed. Some studies choose on the basis of Earth minerals, with typical values around 2-4$~{\rm J} \times {\rm m}^{-1} \times {\rm s}^{-1} \times {\rm K}^{-1}$. For example, \cite{MalamudPrialnik-2015} utilize a greatly simplified mineralogy and use Olivine (anhydrous) and Serpentine (hydrous) for a fully self-consistent derivation of the chemical reaction equations, where the latter minerals have about half the thermal conductivity of the former. Some studies assume particularly small values (e.g., \cite{BlumEtAl-2017} and \cite{GundlachEtAl-2020}) on the basis that they are modeling objects which are very rich in organics, with values as low as 0.5$~{\rm J} \times {\rm m}^{-1} \times {\rm s}^{-1} \times {\rm K}^{-1}$. Another approach is to assume typical meteoritic values, although that approach entails choosing from a huge scope of measurements based on the selection of specimen \citep{OpeilEtAl-2010,OpeilEtAl-2012}, even within the same meteorite class, and could essentially cover anything in the range 0.5-7$~{\rm J} \times {\rm m}^{-1} \times {\rm s}^{-1} \times {\rm K}^{-1}$.
    
    As previously mentioned, since we already have a large set of free parameters, we choose all other parameters to be singly-defined and based on comet 67P/C–G which is the most well-studied cometary archetype. Comet 67P/C–G is extremely rich in organics \citep{BardynEtAl-2017}, with a fraction that might be half of all refractories, hence we adopt a rock thermal conductivity that averages between the \cite{BlumEtAl-2017} organic-rich estimation and the \cite{MalamudPrialnik-2015} pure-mineral estimation. Serpentine and Olivine have thermal conductivities of around 2 and 4$~{\rm J} \times {\rm m}^{-1} \times {\rm s}^{-1} \times {\rm K}^{-1}$, respectively. Hence we adopt $K_{\rm p}=1~{\rm J} \times {\rm m}^{-1} \times {\rm s}^{-1} \times {\rm K}^{-1}$ and $K_{\rm u}=2~{\rm J} \times {\rm m}^{-1} \times {\rm s}^{-1} \times {\rm K}^{-1}$. These values are also reminiscent of values measured in meteorites which might be closer analogues to comets \citep{OpeilEtAl-2010,OpeilEtAl-2012}.
    
    We note that a temperature-independent value seems to be reasonably appropriate beyond about 100 K, for a range of meteoritic materials (for example on the basis of figure 3 in \cite{OpeilEtAl-2012}). We also note that our choices of rock thermal conductivity are considerably larger than those of \cite{BlumEtAl-2017} or \cite{GundlachEtAl-2020}, hence we believe they are extremely conservative upper limits because our thermal evolution simulations, if based on the \cite{BlumEtAl-2017} study, would surely lead to far higher internal temperatures and in turn greater alterations within our simulated comets.
    
    The radiative transport term $K_{\rm rad}$ is also taken from \cite{GundlachEtAl-2020} and references therein. The equation reads

    \begin{equation}\label{EQ:RadiationTerm}
    	K_{\rm rad}(T) \, = \, \frac{16}{3} \, \sigma \, T^3 \, e\, r_{\rm peb} \, (1-\phi) / \phi \ \mathrm{,}
    \end{equation}
    with $\sigma$ being the Stefan-Boltzmann constant, $r_{\rm peb}$ is the pebble radius and $e$ is empirically determined \citep{GundlachBlum-2012}. As mentioned in the beginning of the section, the radiative heat transport is only important when the particle size is large. Hence, when a homogeneous collection of small grains is considered instead of a pile of pebbles, $r_{\rm peb}$ in Equation \ref{EQ:RadiationTerm} must be replaced by $r_{\rm gra}$, and the expression becomes entirely negligible. Likewise, it is easy to show that $K_{\rm rad}$ is also negligible for extremely low temperatures. However, due to the cubic dependence on $T$, it becomes increasingly important for $T>100$ K. $K_{\rm peb}$ becomes subdominant to $K_{\rm rad}$ when $T$ is larger than about 200 K. However, the appearance of liquid water is assumed to destroy the pebble structure, as will be discussed below, and therefore the effects of $K_{\rm rad}$ are limited to no more than the water melting temperature.
    
    The summation of Equations \ref{EQ:net_thermal_conductivity} and \ref{EQ:RadiationTerm} finally gives the effective thermal conductivity:
    
    \begin{equation}\label{EQ:ThermalConductivity}
    	K_{\rm eff}(T) \, = K_{\rm net}(T) + K_{\rm rad}(T) \mathrm{.}
    \end{equation}
    
     We remind that the above description from \cite{GundlachEtAl-2020}, was meant for computing the ejection of pebbles or larger chunks of pebbles. However, in order to model the ejection of the optically dominant sub-pebble dust observed in comets, the simplest but still data-consistent approximation of the pebble structure is in terms of inhomogeneous porous clusters of dust particles \citep{FulleEtAl-2020}, where the dust particles are homogeneous porous agglomerates of sub-micron dust grains \citep{GuttlerEtAl-2019}. In other words, each pebble is in itself a cluster of different-sized sub-pebble agglomerates. Such a structure probably implies even lower thermal conductivity than here assumed, and will be investigated in future models.
    
    \subsubsection{Heat capacity}\label{SSS:HeatCapacity}
    For the heat capacity $c$ of the various water and rock species we generally follow \cite{MalamudPrialnik-2015}. The specific heat capacity of water vapour, denoted by the subscript (v), is given by
    
    \begin{equation}\label{EQ:WaterVaporHeatCapacity}
    	c_{\rm v} \, = 3k_{\rm B} / \mu_{\rm h2o} \mathrm{,}
    \end{equation}
    
    \noindent where $k_{\rm B}$ is the Boltzmann constant and $\mu_{\rm h2o}$ the molecular mass of water. For liquid water we likewise use a constant value, $c_\ell=4186$ J$\times$kg$^{-1}\times$K$^{-1}$. For ice (either amorphous or crystalline) we have a temperature-dependent heat capacity fitted by \cite{Klinger-1980} to data published by \cite{GiauqueStout-1936}:
    
    \begin{equation}\label{EQ:WaterIceHeatCapacity}
    	c_{\rm a,c} \, = 7.49 T +90~\rm{J}\times\rm{kg}^{-1}\times\rm{K}^{-1} \mathrm{.}
    \end{equation}
    
    For rock, the specific heat capacity is likewise temperature-dependent. It is however difficult to provide an exact expression, especially since there are various types of rock. According to the Dulong–Petit law \citep{Kittel-1976}, the molar heat capacity tends to be limited to a single value for all solids at high temperatures because the maximum inner energy in the vibrational mode of the crystal structure is restricted to 3$N_{\rm A}k_{\rm B}$, where $N_{\rm A}$ is the Avogadro constant. However, as the molecular composition becomes more complex, the vibrational energy of the molecules has to be considered, leading to higher heat capacities than implied by the aforementioned limit. To contrast, at low temperatures, below about 100 K, quantum processes become increasingly important. The low-temperature behaviour of specific heat capacity may be described via the Einstein–Debye model \citep{LoehleEtAl-2017}, where it rapidly drops with temperature as $T^3$. Meteorite laboratory measurements generally confirm this behaviour (E.g. in \cite{OpeilEtAl-2012,OpeilEtAl-2020}). Although they do not usually extend beyond room temperature, it is apparent that the specific heat capacity at much larger temperatures tends to approach a particular limit or value. \cite{MalamudPrialnik-2015} adopted an exponential expression to fit both anhydrous rock specific heat capacity (based on \cite{RobieEtAl-1984}) and hydrous rock (based on \cite{BertoldiEtAl-2005}):
    
    \begin{equation}\label{EQ:AnhydrousHeatCapacity}
    	c_{\rm u}(T) \, = 1043 \left( 1 - exp(-3.5\times 10^{-3} T) \right)~\rm{J}\times\rm{kg}^{-1}\times\rm{K}^{-1} \mathrm{,}
    \end{equation}

	\begin{equation}\label{EQ:HydrousHeatCapacity}
		c_{\rm p}(T) \, = 1480 \left( 1 - exp(-3.5\times 10^{-3} T) \right)~\rm{J}\times\rm{kg}^{-1}\times\rm{K}^{-1} \mathrm{.}
	\end{equation}
    
    We find these fits to reasonably capture the behaviour of specific heat capacity in the full range of temperatures. They are also roughly in agreement with meteorite measurements. The expression has a simpler form than e.g. the Einstein model, and it is certainly preferable to constant heat capacity values, which were typically adopted in early studies (see e.g. discussion in \cite{GhoshMcSween-1999}). At high temperatures, the limit heat capacities tend toward a larger value for hydrous compared to anhydrous rocks, as the former have more complex molecules that incorporate hydrogen atoms.
     
    \subsubsection{Pebble compression curve}\label{SSS:PebbleCompression}
    Using laboratory designed experiments, we find that the packing (i.e macro) porosity of pebble piles diminishes from a starting value of around 40\%, when typical pressures exceeding a few kPa are applied. The packing porosity reduces to 0\%, essentially nullifying the original hierarchical pebble structure, for typical pressures in the order of several hundreds of kPa. We thus find that self-gravity can be important for packing porosity compaction when the comet radius exceeds about 10 km. 
    
    The experiments used in order to demonstrate this were first carried out with dust pebbles which were made out of two different materials: polydisperse SiO$_2$ dust, with grains in the diameter range between 0.5 $\mu$m and 10 $\mu$m and a density of $\varrho_{\rm SiO_2}=2600$ kg $\times$ m$^{-3}$; and iron-oxide dust, Fe$_2$O$_3$, with grain diameters less than 5 µm and a density of $\varrho_{\rm Fe_2O_3}=5240$ kg $\times$ m$^{-3}$. The dust pebbles were produced by manually shaking the respective powders, and were initially sieved to obtain only pebbles in the size range 1-1.6 mm. We note that several other dusty materials were experimented with, but only those two materials were successful in both producing and sieving stable dust pebbles. 
    
    In addition to dust pebbles, we performed the same experiment with pure H$_2$O pebbles, in order to obtain an additional compression curve for water ice, the most prevalent volatile inside icy bodies. The H$_2$O ice grains have 1.5 $\mu$m diameters, and a density of $\varrho_{\rm H_2O}=917$ kg $\times$ m$^{-3}$. Their method of production is described in \cite{GundlachEtAl-2011b} and \cite{JostEtAl-2013}. In similarity to the dust pebbles, the H$_2$O powder was manually shaken and subsequently sieved to obtain pebbles in the size range 1-1.6 mm. In order to prevent sintering of the microscopic ice grains inside the pebbles, the temperature in these experiments was kept sufficiently low, according to the description in \citet{Gundlach2018}.
    
\begin{figure}
    	\includegraphics[scale=0.36]{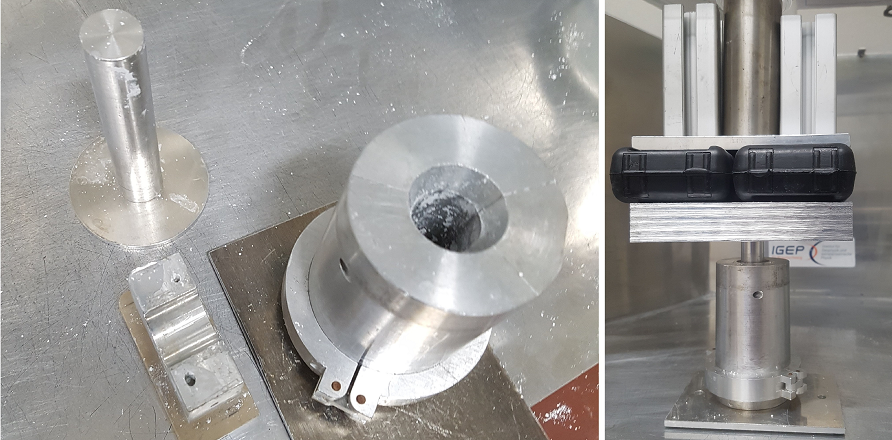}
    	\caption[r]{The left image shows the hollow cylinder and the stamp. On the right the cylinder is shown from the side with the stamp compressing a dust sample using a pressure of 250 kPa.}\label{fig:press}
    \end{figure}

    For the next step, the weight of a certain amount of pebbles was measured and afterwards poured into a hollow cylinder as shown in Figure \ref{fig:press}. Then the pebbles were compressed using a stamp and a weight between 0.1 kg and 8.0 kg. The diameter of the stamp was 1.95 cm and therefore the weight leads to a compaction pressure between 3.3 kPa and 250 kPa. On the right side of Figure \ref{fig:press} a dust sample is compacted with 250 kPa. For the H$_2$O pebbles, the experiments were performed inside a pre-cooled container with a nitrogen atmosphere so as to obtain a low temperature, preventing melting and sintering \citep{Gundlach2018}.

    \begin{figure}
    	\includegraphics[scale=0.4]{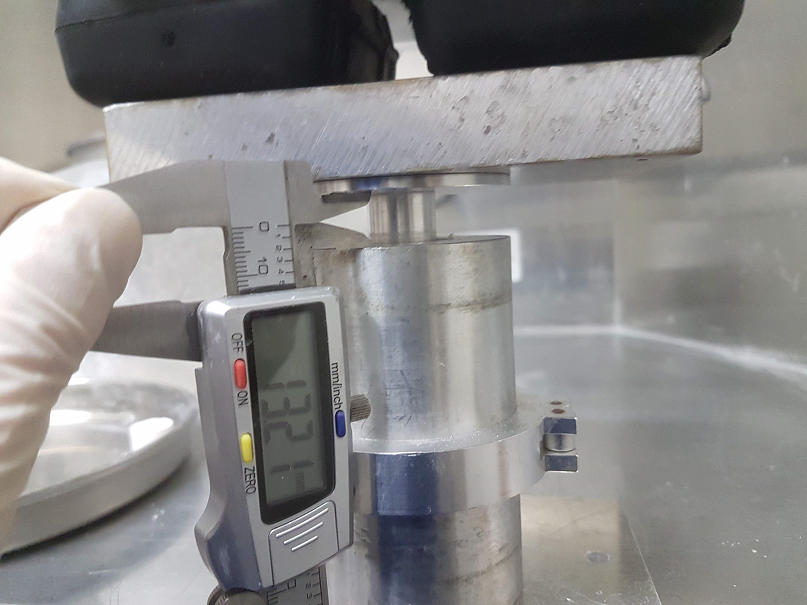}
    	\caption[r]{The height of the dust agglomerate was measured using a digital caliper.}\label{fig:height}
    \end{figure}

    The last step was to measure the height of the compressed dust agglomerate. This was done using a digital caliper while the sample was still compressed as shown in Figure \ref{fig:height}. Once the mass of the agglomerate $m_{\rm agg}$, its radius $r_{\rm agg}$ and height $h_{\rm agg}$ were known, the density $\rho_{agg}$ could be calculated and therefore also the volume filling factor $\phi(p)$, using Equation \ref{EQ:compressionVFF}: 
    
    \begin{equation}\label{EQ:compressionVFF}
    	\phi(p)= \frac{\rho_{\rm agg}(p)}{\varrho}=\frac{m_{\rm agg}}{\pi\cdot r_{\rm agg}^2 \cdot h_{\rm agg}(p) \cdot \varrho}.
    \end{equation}

    After measuring the total volume filling factor, it is possible to further calculate the pebble packing volume filling factor $\phi_{\rm peb}(p)$. Here the volume filling factor of the grains inside the pebbles is assumed to be constant, with approximately $\phi_{\rm gra}=0.4$:
    
    \begin{equation}
    	\phi(p)=\phi_{\rm gra}\cdot\phi_{\rm peb}(p),
    \end{equation}

    \begin{equation}
    	\phi_{\rm peb}(p)=\phi(p)/\phi_{\rm gra} \cong 2.5\cdot \phi(p).
    \end{equation}

    Influence by the Janssen effect cannot be ruled out. The Janssen effect reduces the pressure at the bottom of the sample due to shear forces between the agglomerate and the hollow cylinder \citep{Janssen-1895,Sperl-2005}. The problem can be alleviated by lowering the aspect ratio between the sample and the hollow cylinder. Thus, the experiments using SiO$_2$ were carried out once with samples of 2 g each and once with 4 g each, for greater accuracy.
    
\begin{figure}
    	\includegraphics[scale=0.51]{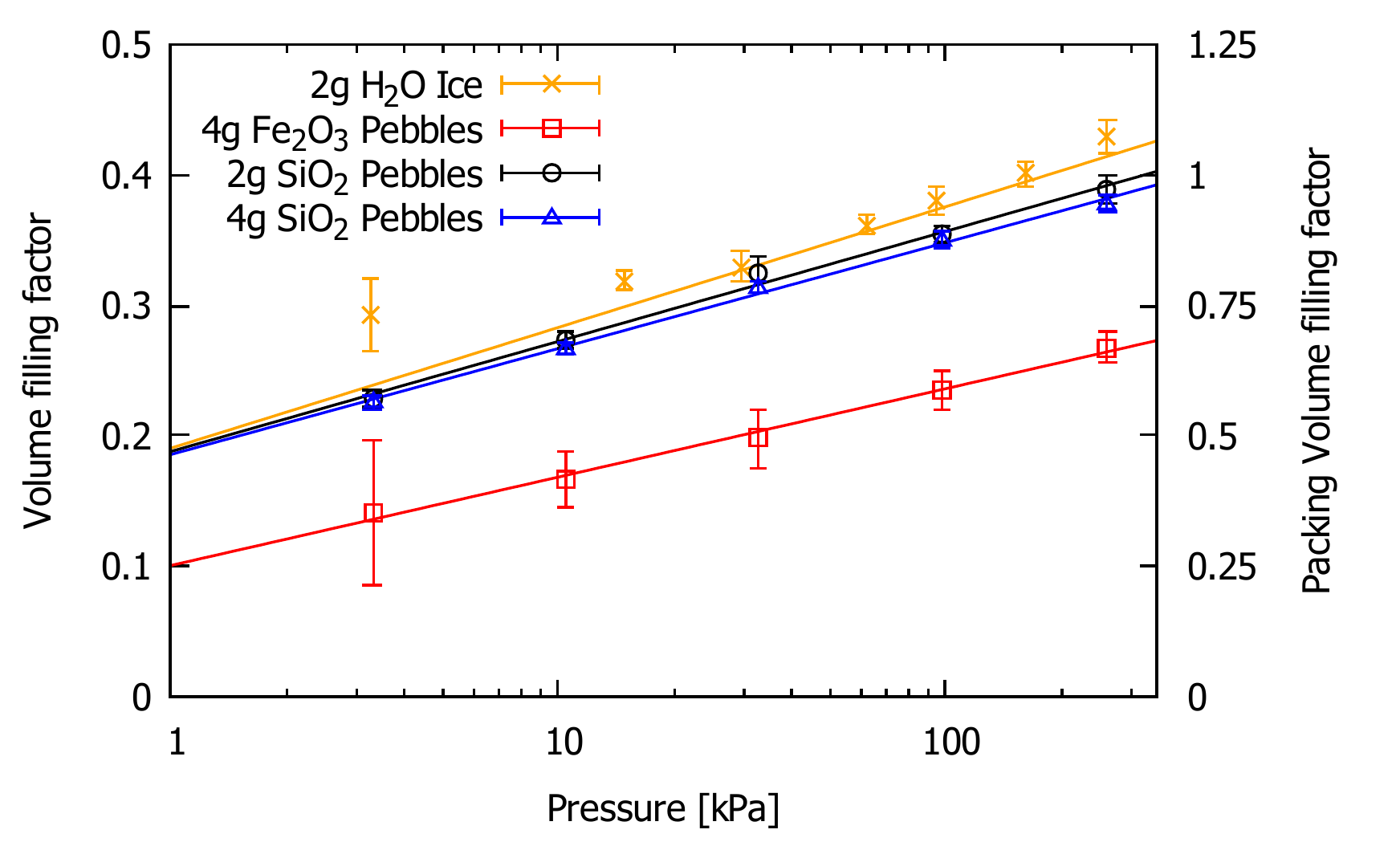}
    	\caption[r]{Shown is the volume filling factor and the packing volume filling factor as a function of the compaction pressure in kPa for SiO$_2$, Fe$_2$O$_3$ and H$_2$O.}\label{fig:VFF}
    \end{figure}

    The results of the experiments are shown in Figure \ref{fig:VFF}. It shows that the packing porosity completely vanishes for the H$_2$O and SiO$_2$ pebbles within the experiment pressure range, while the iron oxide pebbles are less susceptible to packing compression. The H$_2$O pebbles even surpass the point of full packing compression (the micro-porosity of the now homogeneous aggregate starts decreasing). These three materials might not really be appropriate analogues for comet pebbles. The latter are probably composed of a mix of different materials: volatiles, hydrocarbons, various silicates, etc., and have varying material properties, grain sizes and textures. Such details are still not yet fully known, and the complexity of dust chemistry and structure cannot still be properly simulated in laboratory experiments. Given those uncertainties, we assume that the SiO$_2$ pebbles might be more like an 'average' representation of pebbles in cometary setting. Thus, we henceforth apply the SiO$_2$ experiments to our model.
    
    For the SiO$_2$ pebbles the volume filling factor slightly decreases with increasing dust agglomerate mass, as expected (less pressure is transferred onto the cylinder walls). Therefore the results from the experiments with 2 g dust pebbles are used. The fit functions for the 2 g samples shown in Figure \ref{fig:VFF} are:
    
    \begin{equation}\label{EQ:fitVFF}
    	\phi(p)= 0.084 \cdot log(P) +  0.188,
    \end{equation}

    \begin{equation}\label{EQ:fitVFFpack}
    	\phi_{pp}(p)= 0.210 \cdot log(P) + 0.472,
    \end{equation}
    
    \noindent where $P$ is the pressure in units of kPa.
        
\begin{figure}
    	\includegraphics[scale=0.9]{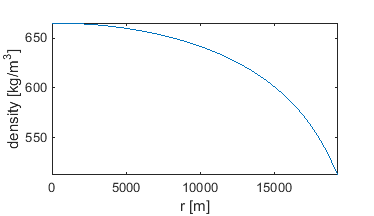}
    	\caption[r]{The density as a function of radial distance from the centre, for a comet whose initial radius is 20 km. After pebble packing compression the radius diminishes only slightly, as the inner parts of the comet compact.}\label{fig:pebble_packing_compression}
    \end{figure}

    From Equation \ref{EQ:fitVFF} we can now easily obtain an equation of state for compressing porous pebble piles, given by the relation $P(\rho)$. We use a similar approach to that described by \cite{MalamudPrialnik-2013} (see their section 2.4). We take the above relation $P(\rho)$, and solve the hydrostatic equation via the methodology given by \cite{PrialnikEtAl-2008} to start our calculation with an initial hydrostatic density profile. Figure \ref{fig:pebble_packing_compression} shows an example of the initial density profile of a large comet whose radius is reduced from 20 km to 19.3 km when pebble packing compression is applied, about 3.5\% reduction. For a larger, 50 km radius comet, the initial radial decrease by packing compression is around 10\%. Below a radius of about 10 km, the reduction is marginal (about 1/2\%) and pebble packing compression by self-gravity may be ignored. Uncompressed, the comet has a uniform density of 533 kg$\times$m$^{-3}$, which becomes, post-compression, approximately equal to the density at the surface of large comets (the surface is where the pressure of self-gravity is always negligible).
    
    However, after setting the initial conditions we do not impose a hydrostatic profile throughout the calculation. Our choice to employ the solution only at the beginning of the calculation originates from the fact that even as the comet warms and volatiles migrate out and freeze in the cold outer layers, while it can change the density profile, the pressure by self-gravity remains similar throughout the calculation. Additionally, neither does vapour transport desiccate the pebbles structure or undermine its stability \citep{HaackEtAl-2021a,HaackEtAl-2021b,SpadacciaEtAl-2021}, as will be further discussed in Section \ref{SS:WaterVapor}.

    We also use Equation \ref{EQ:fitVFFpack} in order to calculate the parameters which govern vapour flow. E.g., equations 13 and 24 in \cite{PrialnikMerk-2008} which calculate the vapour pressure and permeability respectively, now take instead of the overall porosity $\psi$, the packing porosity between the pebbles in which flow is facilitated.
    
    Finally, we assume that the pebbles may collapse completely and instantaneously due to the appearance of liquid water (irrespective of the aforementioned effect by self-gravity). The internal cohesive strength of pebbles is around several kPa. By adding liquid water with a static dielectric permittivity of around 80, we reduce all cohesive forces by a similar factor, i.e. the cohesive strength is reduced to several dozen Pa. Wherever inside the planetesimal the pressure of self-gravity exceeds this low threshold, we assume the pebbles will collapse. We also performed a small experiment, in which we test the hypothesis of aqueous collapse. We produced a stable, spherical silicate pebble, and after wetting the pebble by liquid water it immediately collapsed just by its own weight.
    
    In our model, we collapse the pebbles by adjusting the local porosity to that of a standard aggregate composed of homogeneous micron-sized grains, around 40\% \citep{YuZou-1998}. Since the flow is now channelled through the space between the constituent grains, and not between the pebbles, we change the local pore size to match the new arrangement wherein the spaces between grains are micron-sized, greatly affecting the permeability coefficients and all other parameters that govern physical flow. Heat flow is likewise affected: the structure is no longer hierarchical, so Equations \ref{EQ:net_thermal_conductivity} and \ref{EQ:pebble_thermal_conductivity} are combined with $H_{\rm peb}=1$, and the radiative term is negligible.  
    
    
    \subsubsection{Radiogenic heating}\label{SSS:RadiogenicHeating}
    For calculating the radiogenic heating we require the initial mass fraction $X_0$ of radionuclide isotopes inside the silicate minerals and their associated properties: (a) the half-life $\tau_{\rm h}$ is related to the e-folding time scale $\tau_{\rm e}$ as $\tau_{\rm h}=-\ln (1/2) \tau_{\rm e}$; and (b) the heat released per unit mass, denoted by ${\cal H}$. We compare the canonical values typically used in our model (see e.g. \cite{Prialnik-2002}, table III), to those from the more recent model by \cite{Davidsson-2021} and references therein. We find, like \cite{Davidsson-2021}, that published parameters generally agree within 10\%. We hence use similar values to those from \cite{Prialnik-2002}, except the isotope $^{60}$Fe, for which we use the \cite{Davidsson-2021} estimates. Our parameters are given in Table \ref{tab:Radiogenic}. We note for completion that an additional SLR, $^{53}$Mn, is sometimes also used in thermal evolution models, however as its overall contribution to the heat budget is considered negligible \citep{Castillo-RogezEtAl-2007}, we may justifiably ignore it.
    
    \begin{table}
    	\caption{Radiogenic isotopes and their related parameters at the formation of CAI.}
    	\centering
    	\smallskip
    	\begin{minipage}{14.2cm}
    		\begin{tabular}{|l|l|l|l|}
    			\hline
    			{\bf Isotope} & {\bf $X_0$} & {\bf $\tau_{\rm e}$ (Myr)} & {\bf ${\cal H}$ (J/kg)}\\ 
    			\hline
    			$^{26}$Al & $6.7\times 10^{-7}$  & 1.06  & $1.48\times 10^{13}$\\
    			$^{60}$Fe & $3.46\times 10^{-7}$ & 3.81  & $4.92\times 10^{12}$\\
    			$^{235}$U & $6.16\times 10^{-9}$ & 1030  & $1.86\times 10^{13}$\\
    			$^{40}$K  & $1.13\times 10^{-6}$ & 1822  & $1.72\times 10^{12}$\\
    			$^{238}$U & $2.18\times 10^{-8}$ & 6498  & $1.92\times 10^{13}$\\
    			$^{232}$Th& $5.52\times 10^{-8}$ & 20032 & $1.65\times 10^{13}$\\
    			\hline
    		\end{tabular}
    		\label{tab:Radiogenic}
    		\newline Sources: \citep{Prialnik-2002}; \citep{Davidsson-2021}.
    	\end{minipage}
    \end{table}
    
    \subsection{Caveats and limitations}\label{S:Caveats}   
	The code used in this study is 1-dimensional and relies on the approximation that the object is composed of concentric spherical shells. As comets are usually non-spherical, this approximation is often problematic for performing global evolution calculations. However, it is also an acceptable compromise which is often made in similar studies.
	
	While our code is novel in that it regards many new processes or relations never to be introduced simultaneously in previous comet evolution models, its main limitation is that super- and hyper-volatiles such as CO$_2$, CO and other important species are not considered. The latter are potentially significant as internal heat may drive them to accumulate near the cold surface and eventually moderate the comet's behaviour if and when it becomes active (and also advect heat). Nevertheless, as highly volatile species typically constitute only a small fraction relative to water \citep{AHearnEtAl-2012}, they should have a modest impact on the bulk thermal and physical evolution of the comet. Our goal in this study is to investigate the overall structure and the internal temperatures attained during evolution, and this goal can be achieved reasonably well by considering only the various phases of water (amorphous, crystalline, liquid and gas). Ultimately, the model could benefit from adding super- or even hyper-volatile species, however it remains a task to be explored in future research.
	
	Our study probes a very large, and computationally costly parameter space (see Section \ref{SS:FreeParameters}). A known problem for modellers is that using too many free parameters, it can become exceedingly difficult to comprehend and even present model results. Thus, in this study we decide to explore only one realisation for the initial rock/ice mass ratio, taking after comet 67P/C–G. In the future however, the scope may be broadened, by assuming lower rock/ice ratios (which were more frequently advocated in early comet modelling), or rather assuming larger rock/ice ratios such as 5 or 6, according to some modern estimations \citep{FulleEtAl-2017,LorekEtAl-2018,FulleEtAl-2019,CambianicaEtAl-2020}. This however remains to be investigated in forthcoming studies. 
	
	Finally, We also note that our evolution begins after the optical thinning of the primordial planetary disk and we do not consider any aspects of the evolution prior to that stage.
    
    \section{Parameter study}\label{S:ParameterStudy}
    \subsection{Free parameters}\label{SS:FreeParameters}
    Keeping all other model parameters constant, the free parameters in our study are:
    
    (a) \emph{Initial comet size} - we select comet radii that span the full range of sizes: $R=$ 0.5, 1, 2, 5, 10, 20, 50 km. There are only very few comets that potentially might have radii in the range 20-50 km, including C/2011 KP38, with an estimated radius of 27.5 km \citep{BauerEtAl-2013,IvanovaEtAl-2021}, 29P/Schwassmann–Wachmann with $\sim$30 km \citep{SchambeauEtAl-2015}, Hale-Bopp with 37 km \citep{SzaboEtAl-2012}, C/2002 VQ94 with $\sim$50 km \citep{KorsunEtAl-2014} and only one newly discovered comet C/2014 UN271 (Bernardinelli-Bernstein) which has an estimated radius as large as 70 km \citep{BernardinelliEtAl-2021,HuiEtAl-2022}. Setting aside these outlier objects, most comets with well-known properties actually have radii less than 10 km (see e.g. figure 11 in \cite{GroussinEtAl-2019}). For completion, we also note that in our simulations, comets with a radii in excess of 10 km actually start with a slightly smaller initial radii. The difference comes from pebble packing compression by self-gravity (see Section \ref{SSS:PebbleCompression}), however for simplicity we label their radii with round, pre-compression values.
    
    (b) \emph{Formation time} - radiogenic heating in any planetesimal is related to its formation time after the formation of CAIs, labelled as $t_{\rm CAI}$. The earlier the formation, the larger the abundance of SLRs (the initial mass fraction in Table \ref{tab:Radiogenic} refers to $t=0$ - see also Section \ref{SS:Pristine}). We select the following formation times $t_{\rm CAI}=$ 4, 5, 6, 7, 8, 9, and 10 Myr. This range is motivated by the fact that for values exceeding 10 Myr the radionuclide abundances are already so small that they do not make any significant difference to the evolution. Additionally, young proto-planetary discs usually disperse within that time. To contrast, our investigation shows that below 4 Myr, even the smallest comets might be too thermally processed, as long as most/all refractory materials contribute radionuclides at the level typically inferred from meteorites. The latter assumption is common to all the existing studies reviewed in Section \ref{S:Intro}. However, here we also newly consider the possibility that only an exceedingly small fraction of refractories contribute radionuclides (see below). In that case, the formation times of $t_{\rm CAI}=$ 1, 2 and 3 Myr are considered.
    
    (c) \emph{Pebble size} - we take a binary selection for the pebble radii: 1mm and 1cm. This choice roughly represents the lower and upper limits expected in the literature \citep{ZsomEtAl-2010,LorekEtAl-2018}.    
    
    (d) \emph{Refractory mineral fraction} - radiogenic heating comes from decay of radionuclides (see Table \ref{tab:Radiogenic}), which are preferentially incorporated into the minerals, due to their generally higher condensation temperatures \citep{Lodders2003}. According to \cite{BardynEtAl-2017}, approximately half of the refractories in comet 67P/C–G could be in organics. Hence, potentially, one might argue that only about 50\% of refractories contribute radiogenic material. According to \cite{FulleEtAl-2017}, perhaps comets inherited so little material from the inner Solar system, that as little as 5\% only of all refractories actually contribute radiogenic material. As already mentioned, the fiducial assumption in all the thermophysical studies reviewed in Section \ref{S:Intro} is that 100\% of refractory (so-called 'rocky') materials contribute radionuclides with meteoritic abundances. Here we also consider smaller fractions as an alternative idea. A fraction as small as 0.05 is consistent with the measurements of Al–Mg systematics in four 81P/Wild 2 particles \citep{Levasseur-RegourdEtAl-2018}, where no evidence of live $^{26}$Al at the time of their crystallisation was found. In this case, we assume that the formation time was faster than 4 Myr, because longer formation times would make negligible the radiogenic heating. Thus, we consider a mineral fraction of 0.05 with rapid formation (1-3 Myr), or else mineral fraction of 0.5 or 1 with longer formation times (4-10 Myr).
    
    (e) \emph{Permeability coefficient} - \cite{GundlachEtAl-2020} and references therein introduce a simplifying scaling parameter $b$, representing the number of particle layers required in order to halve the outgassing flux at a given temperature. It can be shown that $b$ is related to the Knudsen diffusivity and in turn the gas permeability coefficient (see \cite{GundlachEtAl-2011a}). These authors find that $b$ (measured in particle diameters) might be in the range 1-7, such that $b=1$ describes the least permeable medium and $b=7$ the most permeable. We take the same approach. Our model calculates the Knudsen diffusivity directly from the temperature and structural parameters, and we may obtain different permeabilities by changing the values of the parameter $\xi$, the tortuosity of the pore space. We take the tortuosity to be in the range 1.4-3.7, which gives approximately the equivalent range of \cite{GundlachEtAl-2020}, i.e $b$ between 1-7. Indeed, recent experiments that focus on the relation between the tortuosity and permeability, measure a similar range in tortuosity (see e.g. figure 19 in \cite{SchweighartEtAl-2021}).
    
    In combination, we have 392 different models for formation times between 4-10 Myr after CAI (Section \ref{SS:4-10Myr}) and 84 different models for formation times between 1-3 Myr after CAI with a single, small mineral fraction (Section \ref{SS:1-3Myr}). Each model is computed on a single CPU, using the Astric I-Core cluster in Israel. The runtime varies between several minutes to weeks, depending on how thermally and physically processed each model evolution gets.
    
    \subsection{Late formation with large mineral fraction}\label{SS:4-10Myr}
\begin{figure*}
    	\includegraphics[scale=0.53]{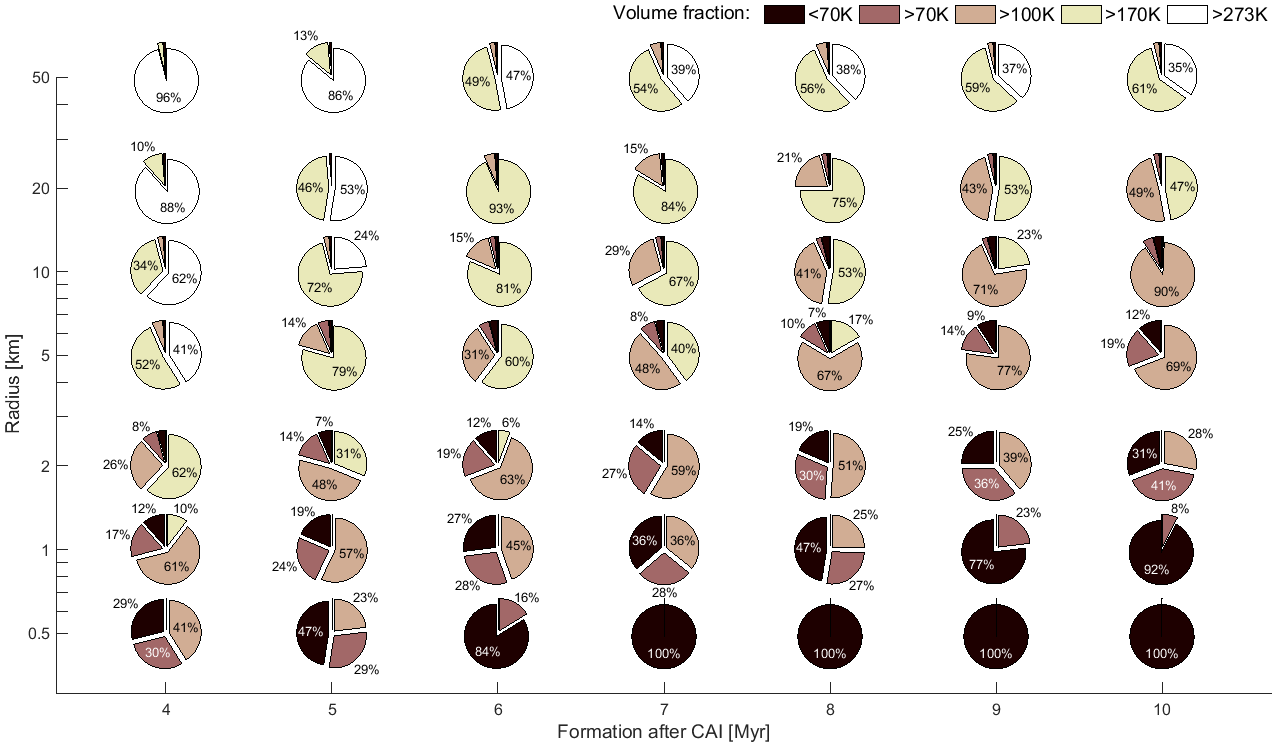}\caption[r]{Pebble radius = 0.1 cm ; Mineral fraction = 0.5 ; Permeability parameter b=1}\label{fig:0.1-0.5-1}
    \end{figure*}

\begin{figure*}
    	\includegraphics[scale=0.53]{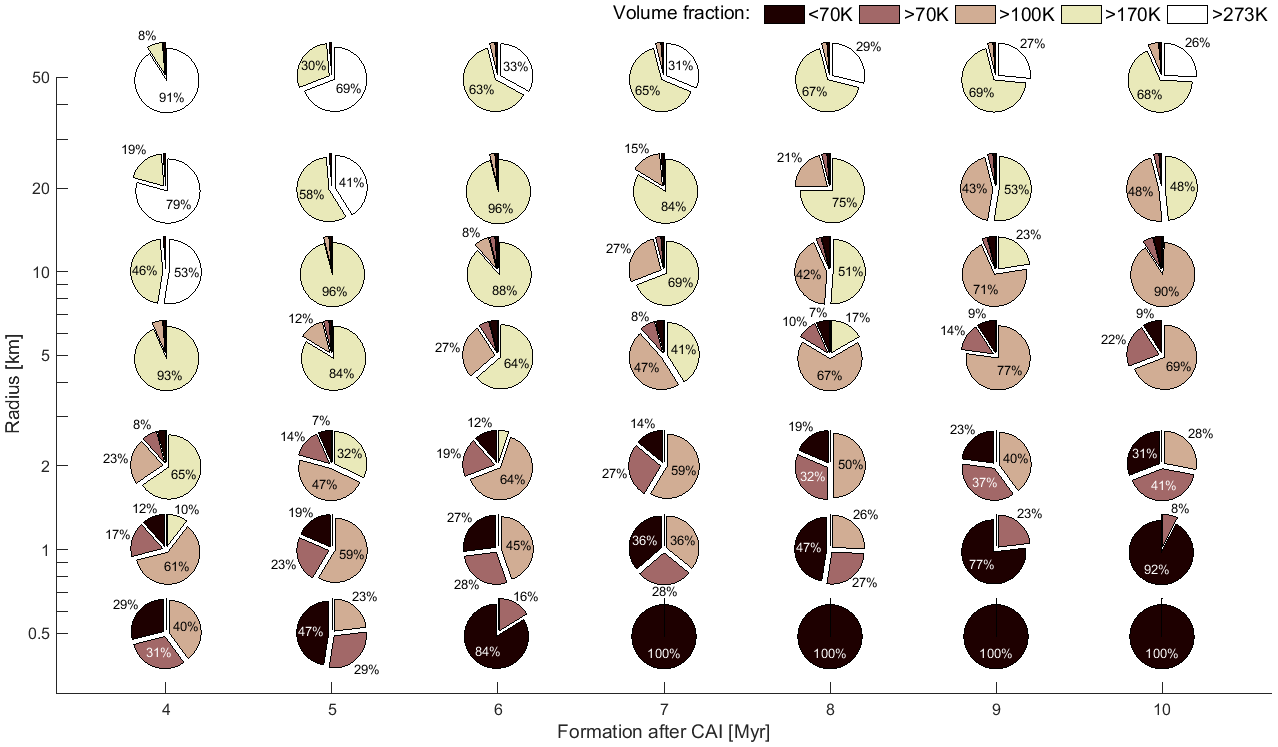}\caption[r]{Pebble radius = 0.1 cm ; Mineral fraction = 0.5 ; Permeability parameter b=7}\label{fig:0.1-0.5-7}
    \end{figure*}

\begin{figure*}
    	\includegraphics[scale=0.53]{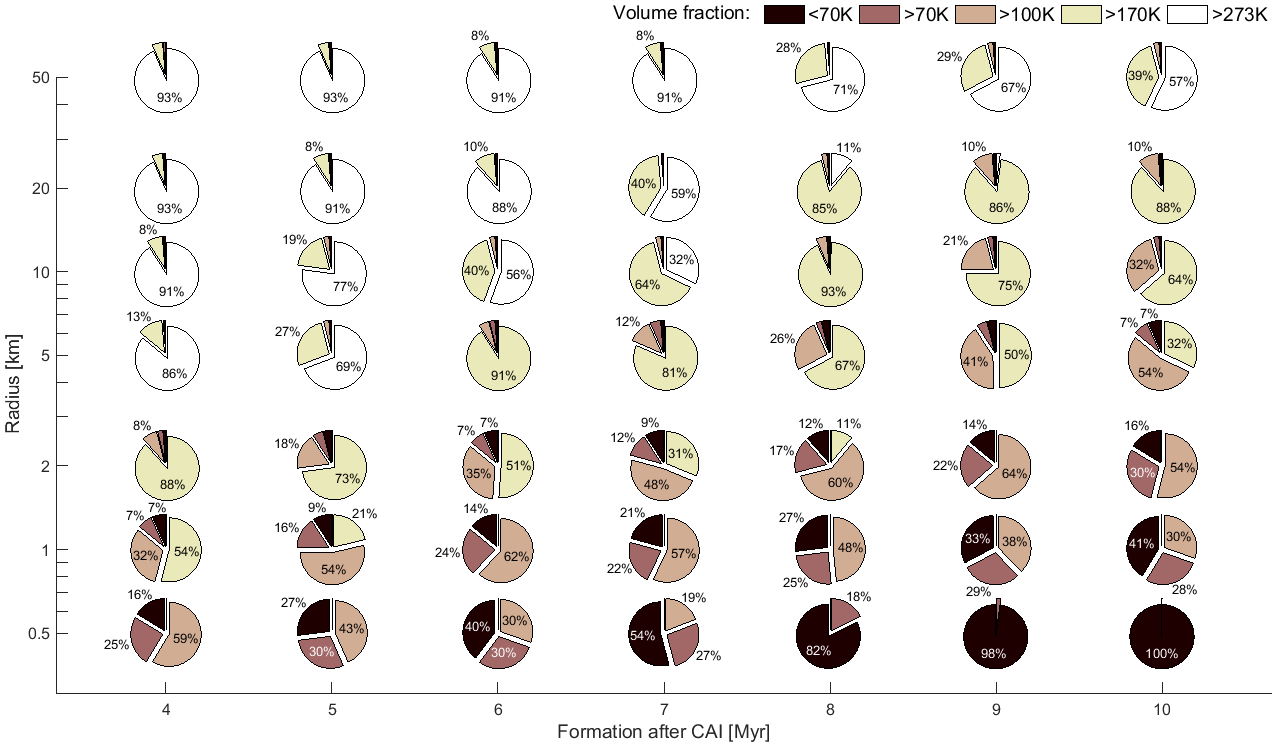}\caption[r]{Pebble radius = 0.1 cm ; Mineral fraction = 1 ; Permeability parameter b=1}\label{fig:0.1-1-1}
    \end{figure*}

\begin{figure*}
    	\includegraphics[scale=0.53]{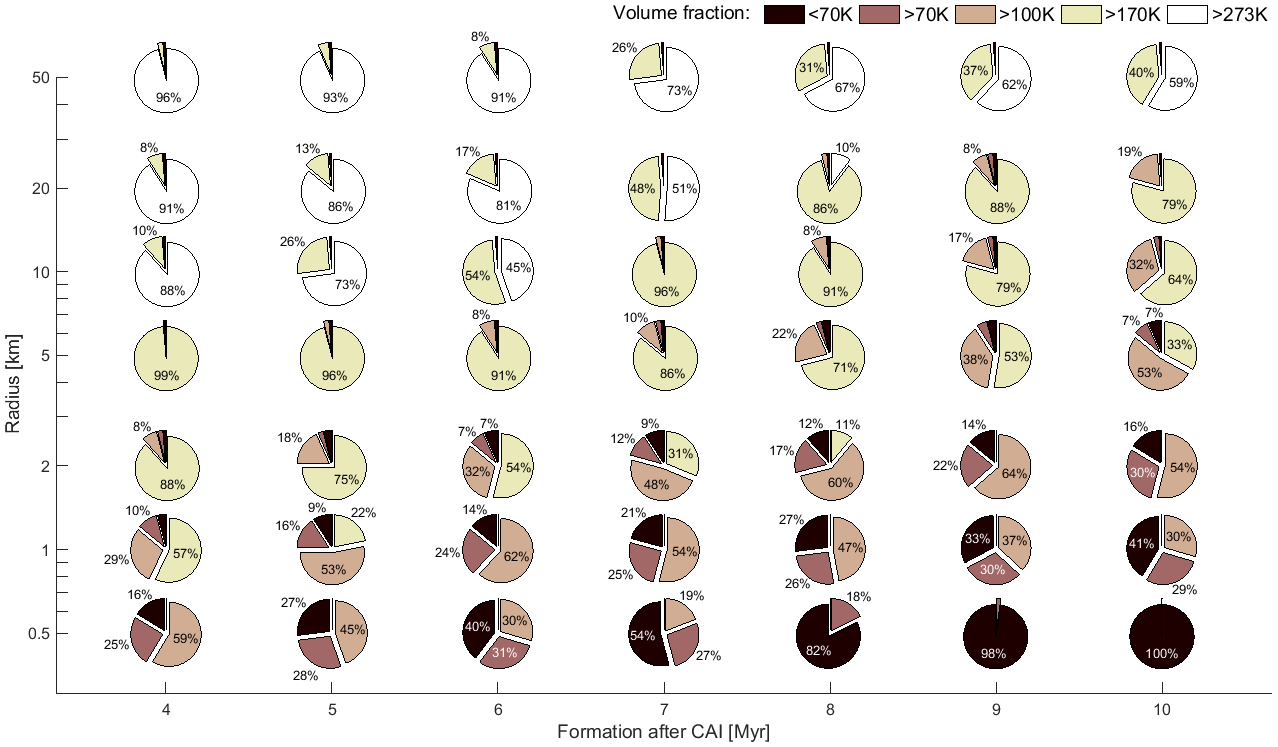}\caption[r]{Pebble radius = 0.1 cm ; Mineral fraction = 1 ; Permeability parameter b=7}\label{fig:0.1-1-7}
    \end{figure*}

\begin{figure*}
    	\includegraphics[scale=0.53]{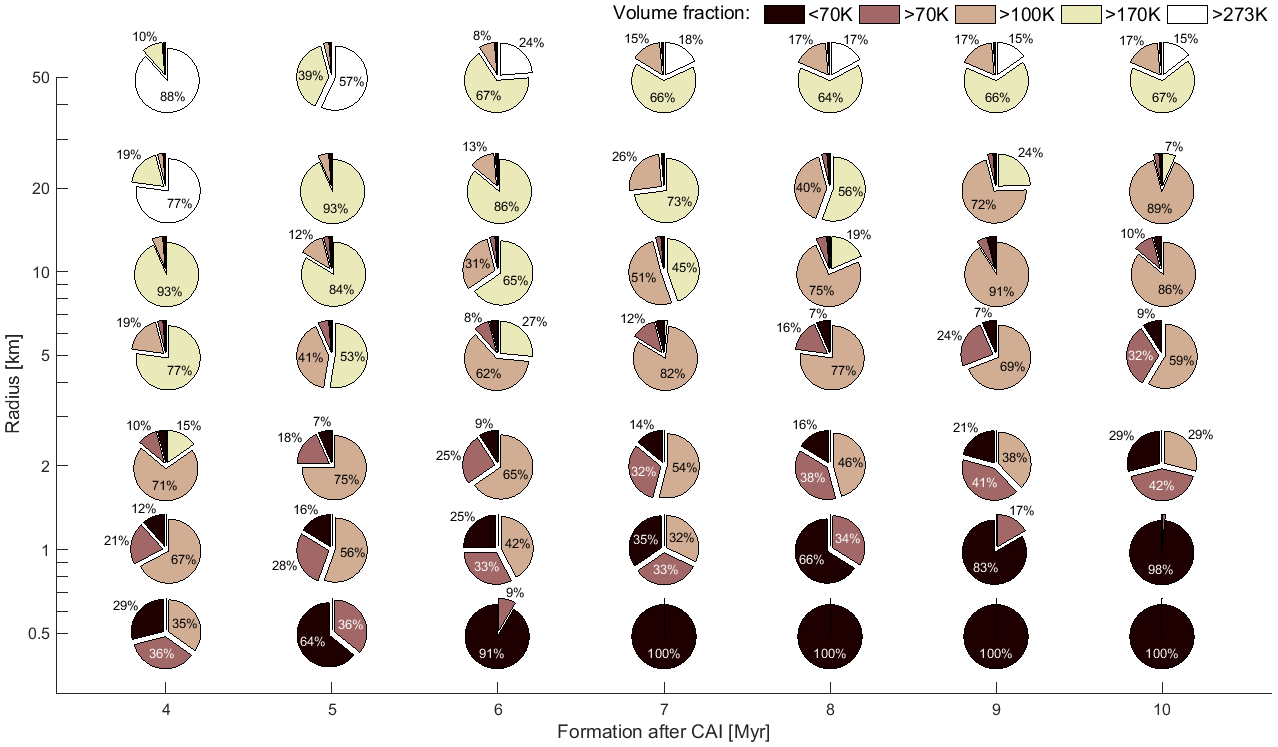}\caption[r]{Pebble radius = 1 cm ; Mineral fraction = 0.5 ; Permeability parameter b=1}\label{fig:1-0.5-1}
	\end{figure*}

\begin{figure*}
    	\includegraphics[scale=0.53]{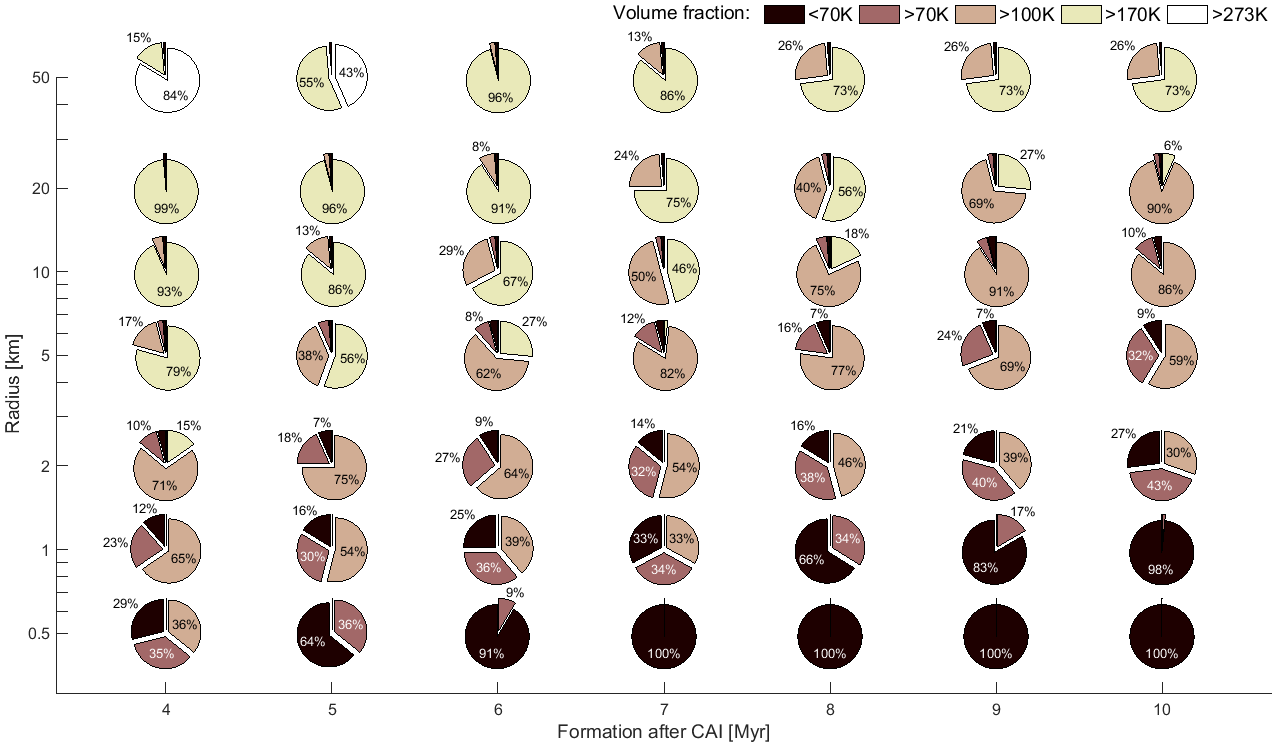}\caption[r]{Pebble radius = 1 cm ; Mineral fraction = 0.5 ; Permeability parameter b=7}\label{fig:1-0.5-7}
	\end{figure*}

\begin{figure*}	
    	\includegraphics[scale=0.53]{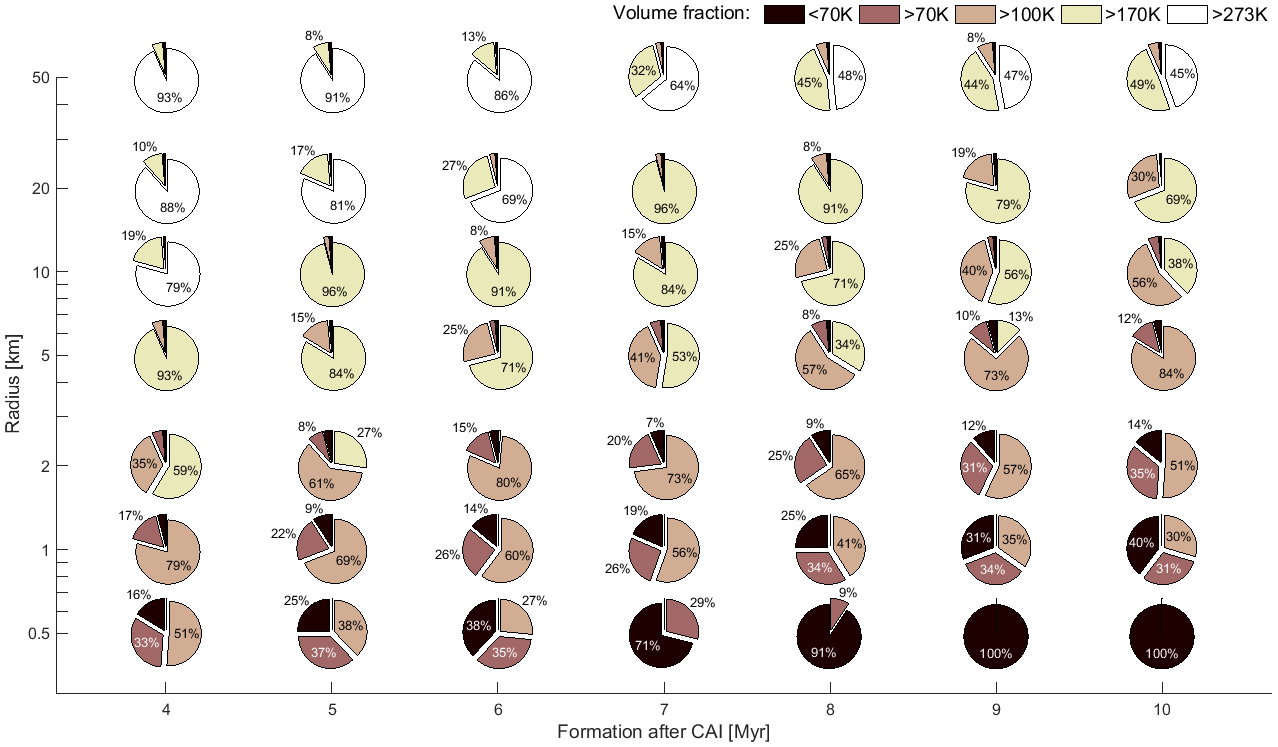}\caption[r]{Pebble radius = 1 cm ; Mineral fraction = 1 ; Permeability parameter b=1}\label{fig:1-1-1}
	\end{figure*}

\begin{figure*}
    	\includegraphics[scale=0.53]{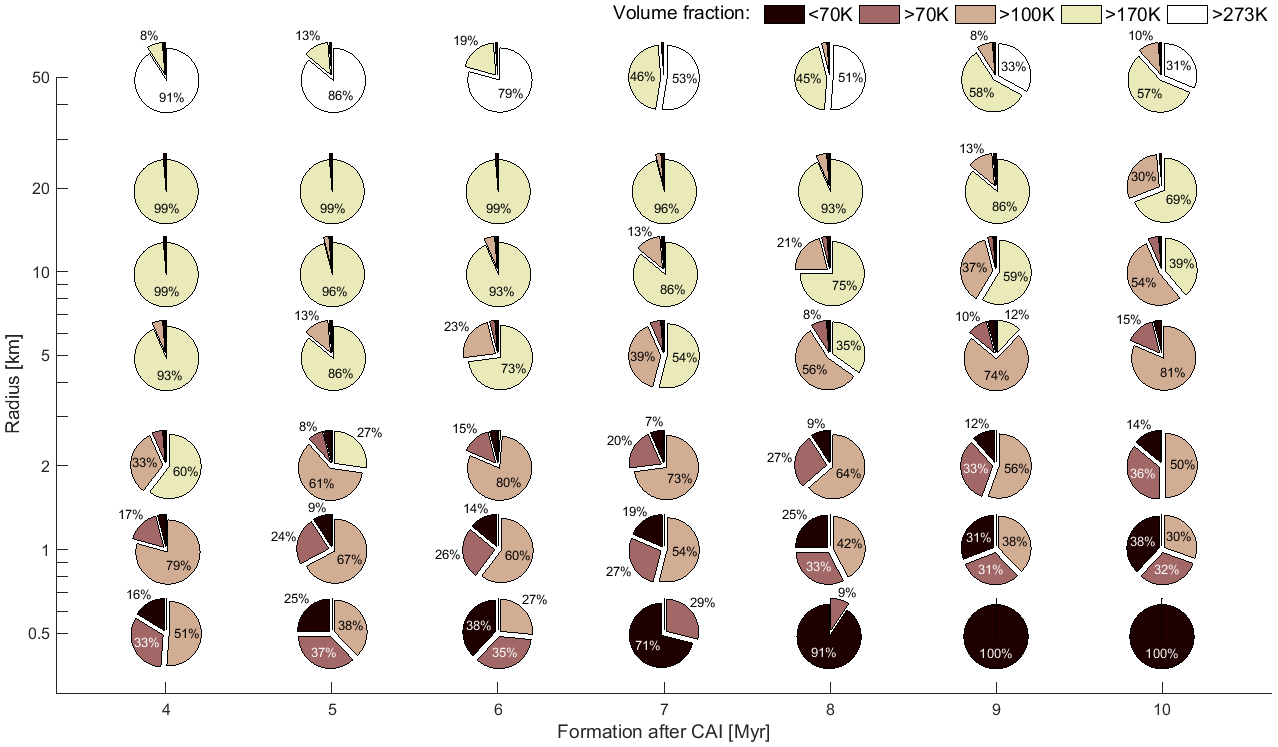}\caption[r]{Pebble radius = 1 cm ; Mineral fraction = 1 ; Permeability parameter b=7}\label{fig:1-1-7}
    \end{figure*}

    Figures \ref{fig:0.1-0.5-1}-\ref{fig:1-1-7} present a compilation of the results from our simulation suite, for the mineral fractions 0.5 and 1, and the various combinations of all other parameters.
    

        \begin{figure*}
		\begin{centering}
			\subfigure[Pebble r$_{\rm peb}$= 0.1 cm ; Permeability b=1]{\label{fig:0.1-0.05-1}\includegraphics[scale=0.67]{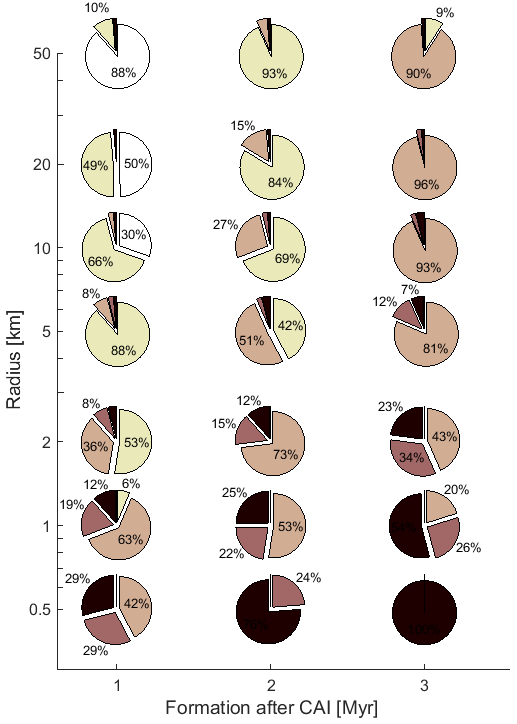}}
			\subfigure[Pebble r$_{\rm peb}$= 0.1 cm ; Permeability b=7]{\label{fig:0.1-0.05-7}\includegraphics[scale=0.67]{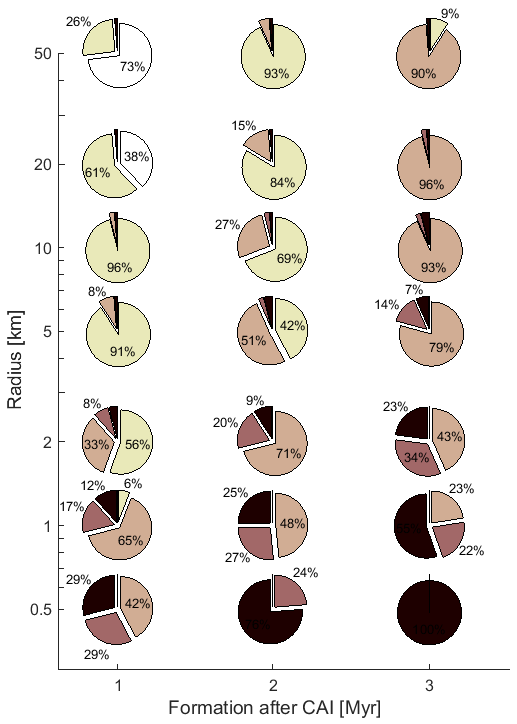}}
			\subfigure[Pebble r$_{\rm peb}$= 1 cm ; Permeability b=1]{\label{fig:1-0.05-1}\includegraphics[scale=0.67]{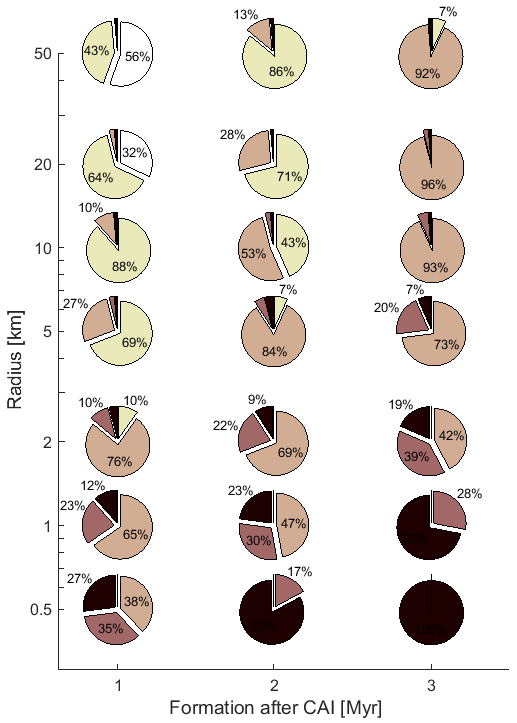}}
			\subfigure[Pebble r$_{\rm peb}$= 1 cm ; Permeability b=7]{\label{fig:1-0.05-7}\includegraphics[scale=0.67]{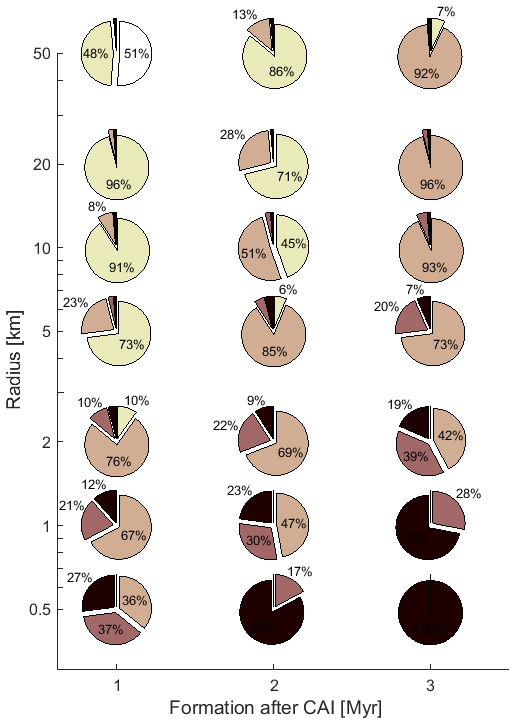}}
			\subfigure{\includegraphics[scale=0.7]{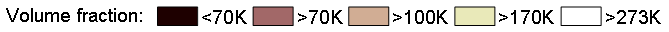}} 
			
		\end{centering}
		\caption{Early formation mosaic: similar to Figures \ref{fig:0.1-0.5-1}-\ref{fig:1-1-7}, but with small mineral fraction of 0.05 only}
		\label{fig:0.05} 
	\end{figure*}
	
    Each simulation outcome is denoted by a single pie chart, that maps the volume fractions within the comet belonging to pre-defined temperature bins. The definition of those bins corresponds to pivotal processes that occur during the evolution: migration of hyper-volatiles only ($<$70 K); sublimation of CO$_2$ and release of trapped gasses therein (70-100 K); phase transitions of amorphous/cubic ice and release of trapped gasses therein (100-170 K); migration of water vapour (170-273 K); aqueous migration and alterations ($>$273 K). We greatly expand on each of these in Section \ref{S:Archetypes}.

    Due to the large parameter space involved in our study, and the considerable number of free parameters, we present each pie chart against the planetesimal radius (y-axis) and formation time (x-axis). The other free parameters, pebble size, mineral fraction and permeability coefficient are binary, and thus we have 8 possible alternative realisations, which we denote in the captions of Figures \ref{fig:0.1-0.5-1}-\ref{fig:1-1-7}.
    
	Figures \ref{fig:0.1-0.5-1}-\ref{fig:1-1-7} trivially show that thermal processing correlates with planetesimal size, and anti-correlates with formation time. In Section \ref{S:Archetypes} we broadly identify general categories of various evolutionary paths, which have distinct implications for the inner-workings and observational consequences of each planetesimal. In Section \ref{S:Discussion} we expand on the significance that these findings might have for the Solar system formation history, and the present day structure of comets.
	
	\subsection{Early formation with small mineral fraction}\label{SS:1-3Myr}
	As mentioned in Section \ref{SS:FreeParameters}-(d), we also consider early formation, which however, is only possible if the mineral fraction is extremely low. Here we use the mineral fraction of 0.05, based on \cite{FulleEtAl-2017} and relying on the isotopic analyses of the only available samples of cometary dust \citep{Levasseur-RegourdEtAl-2018}. Note that a mineral fraction of exactly 1 (following the analysis of meteoritic samples) was assumed in all former models of comet evolution reviewed in Section \ref{S:Intro}. Thus, a mineral fraction of 0.05 represents a major paradigm shift, not only just for comets, but potentially also for other icy objects of similar origin in the outer Solar System.
	
	Nevertheless, early evolution is advocated by some pebble formation models (e.g \cite{PiraniEtAl-2019}) and supported by the presence of dust fractals \citep{FulleBlum-2017}. If applied to our model, the resulting evolutionary outcomes in Figure \ref{fig:0.05} are indeed interesting.
	
	Figure \ref{fig:0.05} is similar to Figures \ref{fig:0.1-0.5-1}-\ref{fig:1-1-7}, although now the mineral fraction is much lower and has a single value. It shows that model outcomes are almost indistinguishable from one another for various choices of the pebble radius and the permeability coefficient. Hence, when the mineral fraction is so small, the most important parameters are the comet's size and formation time. We obtain the result that only large comets ($>\sim$20 km) that formed extremely early ($\sim$1 Myr) are able to reach water melting temperature. There is a clear correlation between the comet size and formation time, and the degree of thermal processing. In sections \ref{S:Archetypes} and \ref{S:Discussion} we elaborate more on the significance of these early formation results.

    \section{Evolutionary archetypes}\label{S:Archetypes}
    One of the main reasons comets are considered interesting, is for typically being thought of as representing a pristine, mostly unaltered population of objects, that have not changed significantly since their natal state in the proto-planetary nebula. This is often justified by the fact that comets are small, and thus are able to cool efficiently. The results in Section \ref{S:ParameterStudy}, however, indicate that it is not necessarily true for most comets, if they are accreted from pebbles. We thereby identify four evolutionary archetypes, which are an attempt to generalise or broadly categorise pebble-accreted comets according to classes, or distinct evolutionary modifications exacted by significantly differing processes. 
    
    \subsection{Pristine}\label{SS:Pristine}
    \noindent \emph{Conditions: extremely small comets, and typically slow to form.}
    \newline
    
    The simplest class of comets is the one in which very few changes occur in terms of the pebble structure or the compositional differentiation inside the comet. We attribute any comet whose peak temperature does not exceed 70 K to this class of objects. At such low temperatures the only certain internal transport is that of hypervolatile gasses such as N$_2$, CO, O$_2$, Ar or CH$_4$, however even these molecules are typically unstable on the cold surface for the comet orbit we assume in this study (see e.g. \cite{LisseEtAl-2021}). If these ices were ever incorporated into comets in pure form (as opposed to impurities released via the processing of other volatile ices), it is easy to imagine that they would be relegated out towards the surface due to internal heating, but hard to imagine how they could be retained near the surface unless the comet has a sufficiently large perihelion distance \citep{LisseEtAl-2021}.
    
    Furthermore, the ratio of all the observed volatile species (primarily CO$_2$) to that of H$_2$O typically does not exceed 20\% \citep{Bockelee-MorvanEtAl-2004,AHearnEtAl-2012}. These estimations are however based on observations of active comet surfaces. Hence, if the surfaces of at least some (or perhaps most) comets concentrate super- or hyper-volatiles by differentiation from the bulk of the comet to merely a small surface layer, the aforementioned ratio is only an upper limit to what may be a much smaller initial value for the bulk of the comet. We must conclude that super- and hyper-volatiles are negligible compared to H$_2$O and are incapable of causing significant changes in the comet's pebble structure, and in turn, modifications to the overall thermal histories investigated in this paper. If anything, small contributions from the sublimation and transport of hyper-volatile ices may only inhibit the progress of thermal evolution, since it consumes and removes radiogenic heat (given that the gas diffusion time scale is such that gas is efficiently removed). Based on these arguments, the fact that hyper- and super-volatiles are not included in our model does not invalidate the main results. 
    
    \begin{figure}   	
    	\subfigure[Temperature]{\includegraphics[scale=0.63]{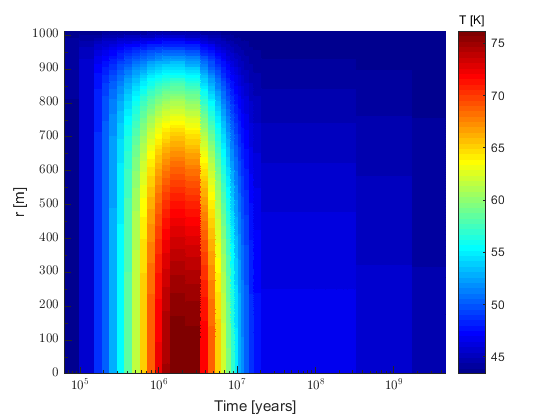}\label{fig:T_1-9-1-0.5-7}}
    	
    	\subfigure[Effective termal conductivity]{\includegraphics[scale=0.63]{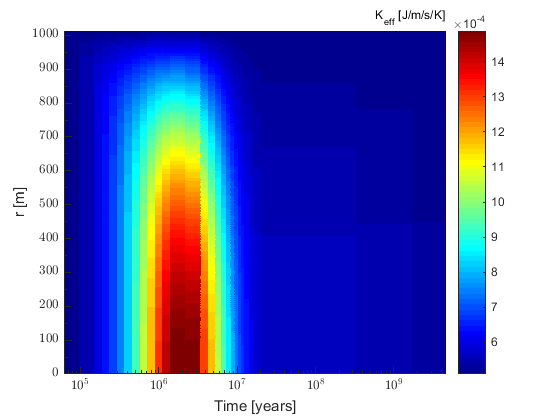}\label{fig:K_1-9-1-0.5-7}}
    	
    	\caption{A 3D plot of the evolving temperature (panel (a)) and effective thermal conductivity (panel (b)), depicted in terms of the radial distance from the centre vs. time, for a 1 km radius comet, with the following parameters: formation time is 9 Myr after the formation of CAI, pebble size is 1 cm, mineral fraction is 1/2 and permeability parameter is $b=7$.}
    	
    	\label{fig:1-9-1-0.5-7}
    \end{figure}

    The results in Section \ref{S:ParameterStudy} indicate that it is next to impossible for comets with radii exceeding 1 km to remain in this category, unless they have an extremely small mineral fraction and, in addition, they form much later than the upper range shown in Figure \ref{fig:0.05} (i.e. much later than 3 Myr). To contrast, sub-kilometre comets are typically sufficiently small in order to remain within this category, but even then the formation time must typically be in the mid-to-upper range of Figures \ref{fig:0.1-0.5-1}-\ref{fig:1-1-7} (large mineral fractions) and \ref{fig:0.05} (small mineral fraction), respectively.
    
    In Figure \ref{fig:1-9-1-0.5-7} we provide one such example, showing the evolution of a 1 km radius comet, similar in size to the small lobe of comet 67P/C–G \citep{SierksEtAl-2015}. We arbitrarily choose the following parameters: formation time is 9 Myr after the formation of CAI, pebble size is 1 cm, mineral fraction is 1/2 and permeability parameter is $b=7$. Panel (a) of Figure \ref{fig:1-9-1-0.5-7} shows the evolution of the temperature, depicted in 3D against the radial distance from the centre of the comet (y-axis) and time (x-axis). Contributions from SLRs are able to increase the temperature from an initial value of 40 K. The peak temperatures are slightly over 70K in the innermost parts, and they are obtained after 2-3 Myr, which corresponds to the SLR time scale. 
    
    Panel (b) of Figure \ref{fig:1-9-1-0.5-7} shows a similar evolution of the effective thermal conductivity $K_{\rm eff}$ (Equation \ref{EQ:ThermalConductivity}). While the grain thermal conductivity is of the order of $2~{\rm J} \times {\rm m}^{-1} \times {\rm s}^{-1} \times {\rm K}^{-1}$ (see $K_{\rm u}$ from Section \ref{SSS:ThermalConductivity}), $K_{\rm eff}$ is more than 3 orders of magnitudes lower. The very low conductivity explains why it is possible for such a small comet to increase in temperature, in opposition to the classical view for objects that size. In the simplest approximation, the behaviour of the temperature comes from the competition between two time scales. As long as the thermal timescale for a given scale length $\Delta r$ is much larger than the radiogenic heating time scale, the heat simply builds up because it cannot escape fast enough. If however the thermal time scale is much smaller than the radiogenic time scale, the comet does not heat up, despite having an internal radioactive heat source. To order of magnitude, the time scale of SLR is around $\sim$1 Myr, and the thermal timescale is given by $\Delta r^2 \cdot \rho \cdot c /(\pi^2 \cdot K)$, where $c$ is the heat capacity, $\rho$ is the density and $K$ the thermal conductivity \citep{HuebnerEtAl-2006}. We can extract $\Delta r$ based on the temperature-averaged thermal conductivity from Panel (b) of Figure \ref{fig:1-9-1-0.5-7} and an approximate value for $c$ from Equation \ref{EQ:AnhydrousHeatCapacity}, to reach a value of about $\sim$400 m. Indeed the smallest objects we consider in our study are of comparable size, which explains why temperature increase is possible.
        
    We also note that since the temperatures are too low for any compositional or structural modifications to occur, the change in the effective thermal conductivity in Panel (b) of Figure \ref{fig:1-9-1-0.5-7} comes only as a result of the increase in the radiative term $K_{\rm rad}$, while $K_{\rm net}$ remains constant. The contribution from $K_{\rm rad}$ is rather significant, since this simulation considers large, 1 cm pebbles.
    
    For comets to remain pristine, the temperature must remain sufficiently low. As previously noted, this generally requires either a small mineral fraction, or at least late formation. For any mineral fraction, late formation inhibits radiogenic heating, which can be easily understood from qualitative arguments. In the limit of infinitely small thermal conductivity, radiogenic energy cannot escape and thus for a slab of mass, the internal heat $u$ is given by the total radiogenic heat and equals the heat capacity times the temperature. $T$ is then approximated via $T=(X_0' \times {\cal H})/c$, where ${\cal H}$ is the energy released per unit mass of radiogenic material (see Table \ref{tab:Radiogenic}) and $X_0'$ is the initial mass fraction of SLR. It can be determined from the formation time of the comet $t_{\rm CAI}$ relative to the e-folding time scale $\tau_{\rm e}$ in Table \ref{tab:Radiogenic}, as $X_0'=X_0 / \exp (t_{\rm CAI}/ \tau_{\rm e})$. Indeed the peak temperatures of all comets in Section \ref{S:ParameterStudy} correlate with the formation time. However, the dependence is much more intricate than in this simple calculation, especially as we consider larger comets in the next sections: LLRs become increasingly important (see Section \ref{SS:WaterVapor}) and the thermal conductivity and heat capacity become increasingly variable (due to extreme changes in temperature and composition).
    
    \subsection{Modifications by occluded super-volatiles}\label{SS:SuperVolatiles}
    \noindent \emph{Conditions: small comets, or intermediate-sized but slow to form.}
    \newline
    
    By contrast to the previous section, super- and hyper-volatiles may be protected inside CO$_2$ or amorphous ice solid matrices as minority impurities. This means that they can be trapped until local temperatures within the comet are sufficient for sublimation of CO$_2$ ice, or indeed the crystallisation of amorphous ice.
    
    The sublimation of CO$_2$ ice typically requires temperatures in excess of 70K (see e.g. characteristic sublimation temperatures in Table 1 of \cite{Yamamoto-1985}). The case for CO$_2$ ice being an important agent for releasing hyper-volatile gasses has recently been strengthened during the Rosseta mission to comet 67P/C–G. It has been argued that the production rates of CO and C$_2$H$_2$ \citep{Luspay-KutiEtAl-2015}, in addition to CH$_4$, HCN, and H$_2$S \citep{GascEtAl-2017}, are all correlated with that of CO$_2$. Laboratory experiments likewise confirm that CO$_2$ is capable of trapping hyper-volatiles \citep{LunaEtAl-2008,SatorreEtAl-2009,SimonEtAl-2019}. This could be consistent with CO$_2$ being the source of these molecules. We briefly note (and discuss in greater detail in Section \ref{S:Discussion}) that if all CO$_2$ in comet 67P/C–G were to migrate from the inner parts of the comet due to internal radiogenic heating, re-condense and concentrate near the surface, it would not have released hyper-volatiles during sublimation, and therefore at least some of its near-surface CO$_2$ must be primordial. This is certainly reasonable as the outer parts of the comet are expected to remain cold prior to becoming active.
    
    In the temperature range between 70-100 K, we hence speculate that (a) super-volatiles such as CO$_2$, NH$_3$, HCN and CH$_3$OH are capable of sublimating and migrating inside the comet \citep{Yamamoto-1985}; and (b) trapped hyper-volatiles might be released from sublimation of CO$_2$ ice. Hyper-volatile materials may be retained in the comet's outer layers only if the comet's orbit warrants sufficiently low surface temperatures.
    
    The phase transition of amorphous ice to crystalline (cubic) ice is typically associated with temperatures in excess of 100 K, where the crystallisation rate starts to become important \citep{SchmittEtAl-1989} and the phase transition releases latent heat, triggering a local increase in temperature. The crystallisation rate is very sensitive to temperature, and will eventually strike a balance with the rate at which heat is conducted, which our simulations indicate typically occurs at around $\sim$110-120 K. 
    
    That amorphous water ice is capable of trapping various hyper-volatiles has been known for many years \citep{Bar-NunEtAl-1985}. Thus, given a similar line of arguments as made previously for CO$_2$, we postulate that in the temperature range of 100-170 K: (a) trapped gasses should be released during the transition from amorphous to cubic water ice; and (b) at somewhat elevated temperatures of around 160 K, additional trapped gasses might be released upon transition from cubic to hexagonal water ice \citep{Davidsson-2021}.        
    
    Unlike in the previous Section, super-volatiles are sufficiently stable given the surface temperatures we consider in this paper, relevant to the Kuiper belt region, and certainly beyond \citep{LisseEtAl-2021}. We would argue that this class of comets, if and when they become active, might display an increased level of activity from super-volatiles since they are transported from the bulk of the comet and are concentrated close to the surface, in addition to pristine super-volatiles already there since the time of formation.
    
    As previously stated, however, our model does not consistently account for the release, flow, or energetic consequences of trapped hyper- or super-volatiles. Our model also ignores the distinction between cubic and hexagonal ice, and the possible release of additional trapped gasses upon transition between those two phases. As already discussed in Section \ref{SS:Pristine}, despite lacking these processes, the thermal evolution outcomes resulting from our current model should not change dramatically even if they were introduced into the code.
    
\begin{figure}
    	{\includegraphics[scale=0.63]{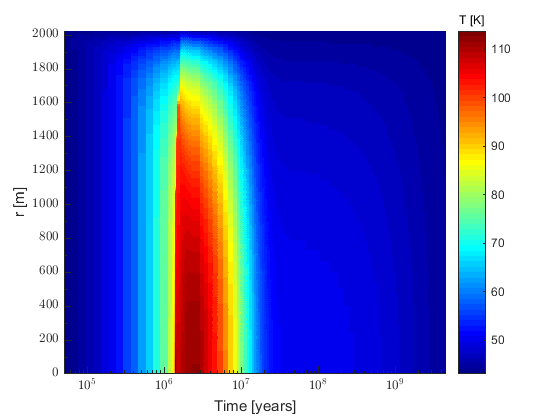}}
    	
    	\caption{A 3D plot of the evolving temperature, depicted in terms of the radial distance from the centre vs. time, for a 2 km radius comet, with the following parameters: formation time is 9 Myr after the formation of CAI, pebble size is 1 cm, mineral fraction is 1/2 and permeability parameter is $b=7$.}
    	
    	\label{fig:2-9-1-0.5-7}
    \end{figure}

    The results in Section \ref{S:ParameterStudy} indicate that most comets whose radii are below 2 kilometres are sufficiently small in order to belong to this category, and in some cases even comet radii in the range 5-10 km, if the formation time is in the upper range of Figures \ref{fig:0.1-0.5-1}-\ref{fig:1-1-7} (large mineral fractions) and \ref{fig:0.05} (small mineral fraction), respectively. In Figure \ref{fig:2-9-1-0.5-7} we provide an example with the same parameter combination to Figure \ref{fig:1-9-1-0.5-7}, only now doubling the comet radius to 2 km, similar e.g to comet 81P/Wild 2 \citep{LiEtAl-2009}. Now the temperature is sufficiently high to trigger the crystallisation of amorphous ice. However, the radiogenic heat, plus the latent heat of crystallisation clearly evident as a spike around 1.4 Myr, are eventually expended. The bulk of the comet cools down to match the surface temperature. Thus, the comet has no thermal memory over the Solar system life time, despite its low thermal conductivity. The top 150 m remain pristine.
    
    \begin{figure}
    	\subfigure[Temperature]{\includegraphics[scale=0.62]{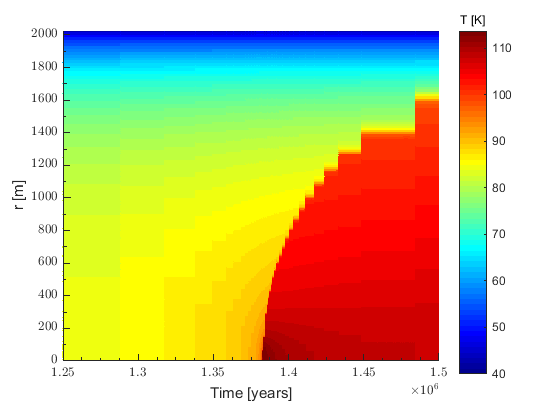}\label{fig:Tzoom_2-9-1-0.5-7}}
    	
    	\subfigure[Amorphous ice density]{\includegraphics[scale=0.62]{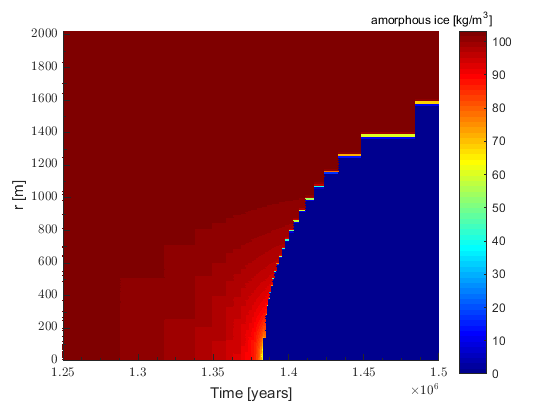}\label{fig:RoAzoom_2-9-1-0.5-7}}
    	
    	\caption{Same as Figure \ref{fig:2-9-1-0.5-7} only zooming-in on the onset of crystallisation, between 1.25 and 1.5 Myr. The evolving temperature (panel (a)) and density of amorphous ice (panel (b)) are depicted in terms of the radial distance vs. time.}
    	
    	\label{fig:zoom2-9-1-0.5-7}
    \end{figure}

    In Figure \ref{fig:zoom2-9-1-0.5-7} we show a zoomed-in segment of Figure \ref{fig:2-9-1-0.5-7} around the time when crystallisation was set off. First in Panel (a) we show the evolution of the temperature, and then in Panel (b) the corresponding evolution of amorphous ice. Given our rock/ice mass ratio of 4, the initial density of amorphous ice is the comet density $\rho=533$ kg $\times$ m$^{-3}$ over (rock/ice+1), or $\sim$107 kg $\times$ m$^{-3}$ (see colour bar).                       
    
    Panel (b) of Figure \ref{fig:zoom2-9-1-0.5-7} shows that the amorphous ice at the centre starts to convert to cubic ice at around 1.3 Myr, however it is not until the temperature reaches 95-100 K, that the rate of crystallisation becomes significant, at which point all the amorphous ice is exhausted at once, releasing latent heat and setting off a chain reaction that propagates out from the central part of the comet. Near the surface, however, the thermal timescale is sufficiently small that amorphous ice remains untransformed.

	\subsection{Modifications by water vapour}\label{SS:WaterVapor}
	\noindent \emph{Conditions: large comets, or intermediate-sized but quick to form.}
	\newline

    The next class of comets would be the first to entail potentially significant compositional changes inside the comet, however the pebble structure remains unaffected by these changes. If the local temperature ever exceeds about 170 K, migration of water vapour starts to become significant. As water is by far the most important and abundant volatile, its transport may significantly changes the internal composition. Vapour differentiation proceeds from the centre outwards. The radial extent to which the comet becomes differentiated is governed by radiogenic heating on the one hand, while energy is being consumed via sublimation and affected by advection, on the other hand. Water migration and eventual re-condensation can change the compositional distribution within, although never in the outermost layers, as long as the surface is cold and inactive.
    
    As already mentioned in Section \ref{SSS:PebbleCompression}, the pebbles are not expected to crumble during the sublimation process, despite losing some of their internal volume. As long as the rock/ice mass fraction is high (which is also the case here, taking after comet 67P/C–G), various studies show that the pebbles do not desiccate when the ice sublimates away \citep{HaackEtAl-2021a,HaackEtAl-2021b,SpadacciaEtAl-2021}. Under certain conditions this might even be true for ice-rich pebbles \citep{SpadacciaEtAl-2021}.
    
    Unlike in the previous two sections, comets belonging to this class of objects are predicted to have cores that could be water-depleted or have diminished water content. Moving from the centre outwards, the water fraction increases due to migration. Closer to the surface, re-condensation enriches the water content and those layers are significantly less porous. Further out, the pristine composition may only become enriched by super- or hyper-volatile ices.
    
    If and when such comets eventually become active, they are expected to display time-varying activity, although arguably such changes will be rather difficult to observe on human time scales. Increasingly, deeper layers are expected to be subject to insolation, and slowly become eroded by sublimation and by ejection of pebbles (as in \cite{GundlachEtAl-2020} or \cite{FulleEtAl-2020}). Depending on the perihelion distance, eroded layers might expel richer deposits of materials in that order: hyper-, then super-volatile ices, then water-rich deposits lacking any super-volatiles, then increasingly less water. If finally such a comet expels all of its volatile-rich layers, we may end up with a fully or partially dry and inert core. That is, in the absence of sufficient volatiles, pressure build-up will be inhibited and pebbles will no longer be able to eject, leading to the formation of an insulating pure dust mantle. Such objects would be largely indistinguishable from most asteroids, perhaps with the exception of rubble pile asteroids, which sometimes lack subcentimeter particles altogether \citep{LaurettaEtAl-2019,CambioniEtAl-2021}. For further discussion see Section \ref{SS:inert}.
    
\begin{figure}
    	\subfigure[Water density]{\includegraphics[scale=0.9]{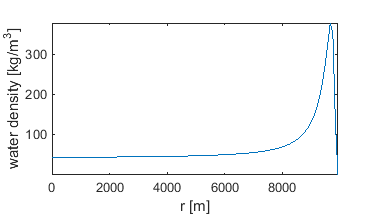}\label{fig:RoWfinal_10-5-1-1-7}}
    	
    	\subfigure[Porosity]{\includegraphics[scale=0.9]{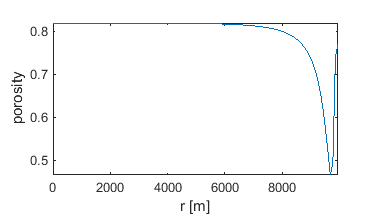}\label{fig:PSIfinal_10-5-1-1-7}}
    	
    	\caption{Following a 4.6 Gyr evolution (final state) of a 10 km radius comet, with the following parameters: formation time is 5 Myr after the formation of CAI, pebble size is 1 cm, mineral fraction is 1 and permeability parameter is $b=7$. Panel (a) shows the water density and Panel (b) the porosity, both as a function of radial distance from the comet centre.}
    	
    	\label{fig:final10-5-1-1-7}
    \end{figure}

\begin{figure}
		\subfigure[Temperature]{\includegraphics[scale=0.62]{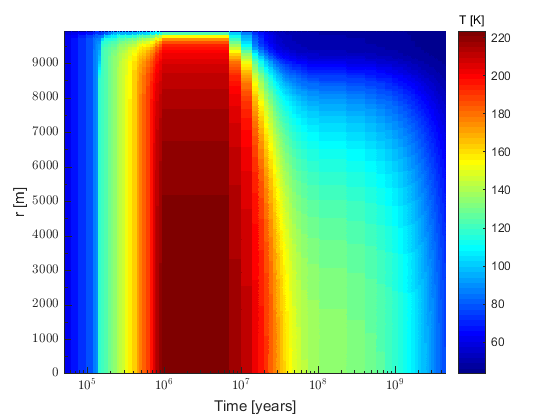}\label{fig:T_10-5-1-1-7}}
		
		\subfigure[Effective thermal conductivity]{\includegraphics[scale=0.62]{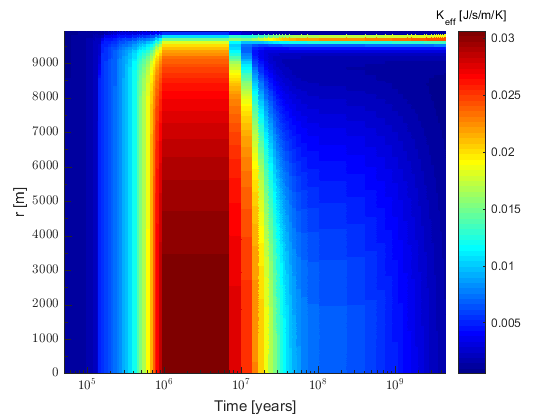}\label{fig:K_10-5-1-1-7}}
		
		\caption{Same as Figure \ref{fig:1-9-1-0.5-7}, only for the comet parameters as in Figure \ref{fig:final10-5-1-1-7}. The evolving temperature is shown in panel (a) and the effective thermal conductivity in panel (b), as a function of radial distance from the centre and time.}
		
		\label{fig:10-5-1-1-7}
	\end{figure}

    We however illustrate by Figures \ref{fig:final10-5-1-1-7} and \ref{fig:10-5-1-1-7} that fully depleting a cometary core of water merely through vapour transport is not trivial, because it requires sufficiently high temperatures that are also maintained for enough time to transport the vapour. This combination is not easy to obtain, as follows. In Figure \ref{fig:final10-5-1-1-7} we show the water density (Panel (a)) and porosity (Panel (b)) after a 4.6 Gyr evolution of a comet with a 10 km radius, a formation time of 5 Myr after the formation of CAI, pebble radius of 1 cm, mineral fraction of 1 and permeability parameter $b=7$. At $t=0$, the water density is 107 kg $\times$ m$^{-3}$ (as already explained in Section \ref{SS:SuperVolatiles}). Panel (a) of Figure \ref{fig:final10-5-1-1-7} shows that vapour transfer has reduced the central water density to less than half of its initial value, while the near surface layers were enriched by up to a factor of 4. Around 250 m below the surface, the rock/ice mass ratio changed from initially 4:1, to around 1:1, considerably reducing the porosity (see Panel (b) of Figure \ref{fig:final10-5-1-1-7}).
    
    Figure \ref{fig:10-5-1-1-7} shows how the distribution of water has reached its current state after only a few $10^7$ yr. The temperature in Panel (a) is peaked after around 3 Myr. It sufficiently exceeds the sublimation temperature of water, hence vapour starts to relegate out from the centre. However, as the energy contribution from SLRs subsides, these high temperatures are not maintained for very long and drop below the necessary threshold in merely a few $10^7$ yr. Meanwhile, Panel (b) shows the corresponding evolution of the effective thermal conductivity. Its behaviour is initially similar to the more pristine comet shown in Panel (b) of Figure \ref{fig:1-9-1-0.5-7}, only here the contribution of the radiative term $K_{\rm rad}$ is more significant than any compositional or porosity changes, owing to the much higher temperatures, so the effective thermal conductivity is around 20 times higher in comparison to the former. After the comet cools down, the effective thermal conductivity drops accordingly. However, $K_{\rm eff}$ remains high in the low porosity layers near the surface. In those layers $K_{\rm net}$ clearly dominates over $K_{\rm eff}$, as re-condensed vapour fills the pore space, affecting the porosity corrections introduced in Equations \ref{EQ:net_thermal_conductivity}-\ref{EQ:ThermalConductivity}.
        
    We have seen that significant vapour transport implies a short formation time, as peak temperatures are determined early-on by SLRs (see Section \ref{SS:Pristine}). We can qualitatively show that LLRs are important only in comets with radii that considerably exceed 20 km, using similar arguments to those already made in Section \ref{SS:Pristine}. To order of magnitude, the time scale of LLRs is around $\sim$1 Gyr, and the thermal timescale is again given by $\Delta r^2 \cdot \rho \cdot c /(\pi^2 \cdot K)$. Approximating $c$ from Equation \ref{EQ:AnhydrousHeatCapacity} and $K$ from Panel (b) of Figure \ref{fig:10-5-1-1-7}, $\Delta r$ is in the range $\sim$40-50 km. Hence, the thermal timescale dictates the course of the evolution and for a 20 km radius LLRs are, at most, able to slow down the cooling, as seen in Panel (a) of Figure \ref{fig:10-5-1-1-7}. Only for comets whose radii exceed 40-50 km, might we expect a second, LLR peak in the temperature, as indeed shown in Figure \ref{fig:T_50-6-1-0.5-7}. However, a second LLR peak in temperature is far from guaranteed. E.g, the value of $K$ may easily increase during an aqueous early evolution, reducing the aformentioned thermal timescale.

	\subsection{Aqueous alterations}\label{SS:AqueousAlteration}
	\noindent \emph{Conditions: extremely large comets, or large comets that form very quickly.}
	\newline

	The final class of comets is the most intricate in terms of its internal processes and their consequences. Unlike all previous classes, the appearance of liquid water not only drastically changes the internal differentiation, due to much more rapid liquid water flow than is possible with merely vapour transport, but also for the first time it may collapse the hierarchical structure of the pebbles, thus far assumed, and considerably alter the rocky composition, with far reaching implications.
	
	As portrayed in the equations in Sections \ref{SSS:ThermalConductivity} and \ref{SSS:HeatCapacity}, the thermal properties of rocky minerals change as a result of aqueous transformation. The specific densities of hydrated minerals are likewise lower than their precursor anhydrous minerals, hence porosity decreases (see e.g. model of Enceladus by \cite{MalamudPrialnik-2013}). The water that participates in the reaction is embedded onto the transformed rock, which changes the net rock/ice mass ratio locally and eventually globally as large internal portions are transformed. Compositional changes due to water absorption are also manifested as thermal and structural changes. The most important contribution of aqueous reactions is the energy that it releases, which contributes vast amounts of heat over a relatively short period of time. This energy serves as a powerful internal heat source. Once triggered, a serpentinization reaction front sweeps through the body. The rate at which it advances depends on the kinematics of flow, so it is both governed by our flow model, while also introducing many feedbacks into the model \citep{MalamudPrialnik-2013}.
	
	Unlike in previous uses of our code \citep{MalamudPrialnik-2013,MalamudPrialnik-2015,MalamudPrialnik-2016,MalamudEtAl-2017}, here the modeling of aqueous processes is slightly more complex, since the material assumed is not merely that of a porous aggregate, as before, but a hierarchical structure of pebbles. Now liquid water is assumed to instantaneously collapse the pebbles, to form a homogeneous aggregate of the pebble's constituent micron-sized grains. In addition to the feedback mentioned in the previous paragraph, we therefore also add the effect of considerably lowering the pore size (by 3-4 orders of magnitude), which strongly affects the permeability of flow. Below we show how all of these effects combined, lead to complete and rapid differentiation, as well as significant change in the global size of the comet, due to pebble collapse.
		
	Aqueous alteration experiments on anhydrous cometary analogue materials \citep{NelsonEtAl-1987,RietmeijerEtAl-2004,Nakamura-MessengerEtAl-2011} have generally confirmed the rapid formation of hydrated phyllosilicate minerals, highlighting how such transitions can occur within mere hours in some cases, and also the significance of the aforementioned changes that may come about, including for example the swelling of hydrated phyllosilicates (having lower specific densities), thus changing the porosity and in turn water permeability.
	
	Although there is no direct and clear evidence for aqueous alterations in comets, various studies do indeed support similar alterations in somewhat related, chondritic meteorites \citep{TakirEtAl-2013,LeGuillouEtAl-2014,JacquetEtAl-2016,LindgrenEtAl-2017,RayEtAl-2021,SuttleEtAl-2021}.
	
\begin{figure}
		{\includegraphics[scale=0.507]{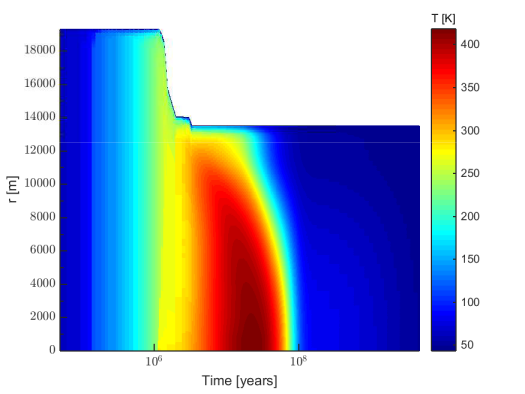}}
		
		\caption{Evolving temperature, depicted in terms of the radial distance from the centre vs. time, for a 20 km radius comet, with the following parameters: formation time is 5 Myr after the formation of CAI, pebble size is 0.1 cm, mineral fraction is 1 and permeability parameter is $b=1$.}
		
		\label{fig:T_20-5-0.1-1-1}
	\end{figure}

	As an example of these processes, we select a simulation of a comet with a 20 km radius and the following parameters: formation time is 5 Myr after the formation of CAI, pebble size is 0.1 cm, mineral fraction is 1 and permeability parameter is $b=1$. In Figure \ref{fig:T_20-5-0.1-1-1} we show the evolution of temperature. From initially $T=40$ K, the temperature increases until eventually the central temperature is sufficiently high to trigger the melting of water, collapse of pebbles and start of a serpentinization chain reaction that sweeps through the comet. Heat from serpentinization increases the temperature even further. The global radius is seen to collapse from initially 19.3 km (just under 20 km, as some pebble packing pre-compression was already in place just by self-gravity). The final radius is merely 13.75 km, an additional 29\% reduction over the initial value. The peak central temperatures are reached at around a few $10^7$ years, after which the comet starts to cool. This is unsurprising as the thermal time scale, $\Delta r^2 \cdot \rho \cdot c /(\pi^2 \cdot K)$, corresponds exactly to a few $10^7$ years, taking $\Delta r=20$ km and assigning typical (averaged) values to the other parameters. After 4.6 Gyr, the central temperature has decreased to slightly over the equilibrium surface temperature, so the comet hardly retains any residual heat at present.
	
\begin{figure}
	    
		\subfigure[Porosity]{\includegraphics[scale=0.63]{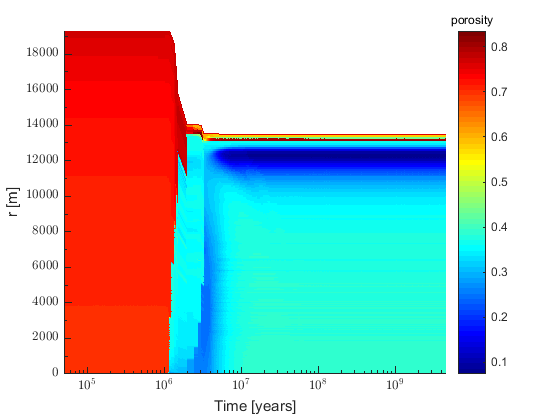}\label{fig:PSI_20-5-0.1-1-1}}
		
		\subfigure[Bulk density]{\includegraphics[scale=0.63]{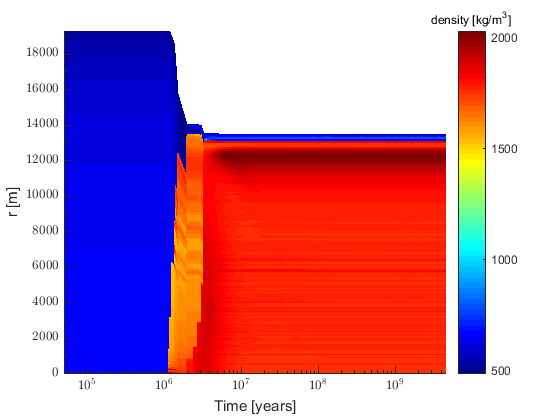}\label{fig:RHO_20-5-1-1-1}}
		
		\caption{Same as Figure \ref{fig:T_20-5-0.1-1-1}, however here the evolving porosity is shown in panel (a) and bulk (total) density is shown in panel (b), as a function of radial distance from the centre and time.}
		\label{fig:20-5-0.1-1-1}

	\end{figure}

	In Figure \ref{fig:20-5-0.1-1-1} we show how the comet structure evolves. Panel (a) shows the evolution of porosity. It can be seen that the initial porosity is not entirely uniform, ranging from 77\% at the surface to 70\% at the centre. This comes from pebble pre-compression by self-gravity, as was also shown in Figure \ref{fig:pebble_packing_compression} in terms of density. The first major change occurs after the onset of water melting and pebble collapse. In the inner parts of the comet, the porosity among the solid grains drops to about 40\% (see Section \ref{SSS:PebbleCompression}), as indicated by the light blue tones. However, some of the pore space is filled by liquid water, making the porosity lower than 40\%, as indicated by the darker blue tones. As the liquid water flows out, the dark blue tones gradually transition to the lighter ones. Liquid water migration however causes internal differentiation inside the comet, and where liquid water eventually refreezes, the porosity closes and even drops below 10\%. A vertical porosity gradient is easily seen. The porosity profile stops evolving after a few $10^7$ years, around the time the comet begins to cool.
	
	In the outer part (top kilometre) of the comet, temperatures are sufficiently high to drive water vapour transport from the base of this layer towards the surface, through the still in-tact pebbles. A similar vertical porosity gradient near the surface therefore indicates the re-condensation of vapour, enriching those layers with water ice. At the base of this layer lies a water-free pebble layer. Its porosity is higher than the value it had prior to vapour transport, and reaches up to 88\%, because the specific density of water ice grains is indeed much lower than that of rocky grains.
	
	The corresponding density evolution is shown in Panel (b) of Figure \ref{fig:20-5-0.1-1-1}. After pebble collapse and serpentinization of the rock, the expected density is ($\phi \cdot \varrho_{\rm u}$), i.e. the volume filling factor $\phi=$(1-$\psi$) times the specific density of hydrated rock. From Section \ref{SSS:ThermalConductivity} we have $(1-0.4) \times 2900$ kg $\times$ m$^{-3} = 1740$ kg $\times$ m$^{-3}$. That is precisely the density indicated by the red tones, except where the pore space is filled by additional frozen ice, in which case the density is slightly increased, reaching up to 2000 kg $\times$ m$^{-3}$. In the outer parts of the comet, the density remains similar to before, since no aqueous pebble collapse has occurred.

	\begin{figure}
		\subfigure[Cumulative energy released]{\includegraphics[scale=0.62]{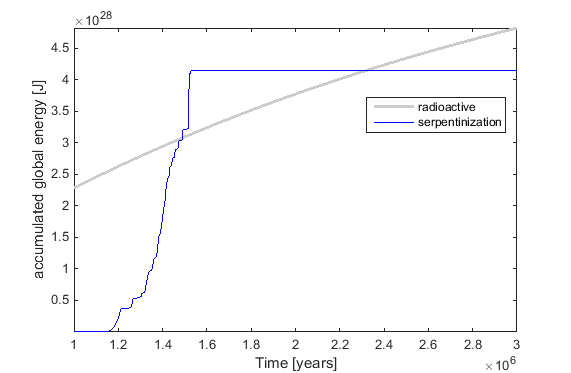}\label{fig:E_20-5-0.1-1-1}}
		
		\subfigure[Rock/ice mass ratio]{\includegraphics[scale=0.62]{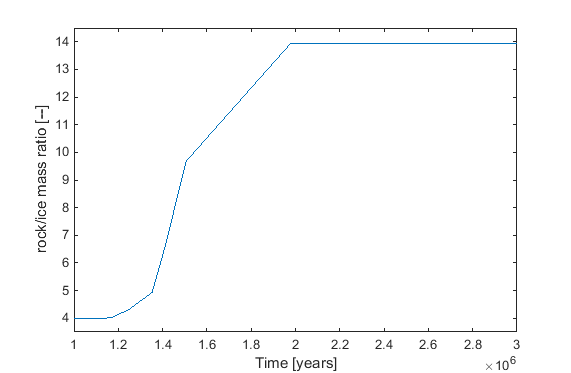}\label{fig:Rocktoice_20-5-0.1-1-1}}
		
		\caption{The early evolution of the comet from Figure \ref{fig:T_20-5-0.1-1-1} between 1 and 3 Myr. Panel (a) shows the cumulative energy released from radiogenic heating (thick grey line) and heat of serpentinization (thin blue line) and Panel (b) shows the corresponding global rock/ice mass ratio inside the comet, evolving due to serpentinization and water absorption.}
		
		\label{fig:early20-5-0.1-1-1}
	\end{figure}

\begin{figure}
		{\includegraphics[scale=0.9]{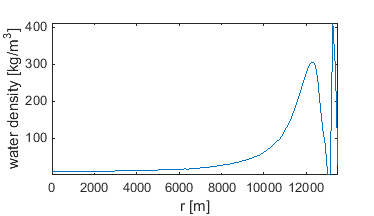}}
	
		\caption{The density profile of water ice after 4.6 Gyr of evolution of a 20 km radius comet, with the parameters from Figure \ref{fig:T_20-5-0.1-1-1}.}
	
		\label{fig:RoWfinal_20-5-0.1-1-1}
	\end{figure}

	Figure \ref{fig:early20-5-0.1-1-1} illustrates the important consequences of serpentinization, zooming-in to the relevant time interval between 1 and 3 Myr. Panel (a) shows the cumulative energy released from both radiogenic heating (thick grey line) and serpentinization (thin blue line). While radiogenic heating is more important throughout the evolution, serpentinization contributes a comparable amount of energy during the critical, early evolution. More importantly, once serpentinization is triggered, it initially releases its energy at a rate two and occasionally even three orders of magnitudes higher than radiogenic heating. The serpentinization energy is so quick to be released, that it actually surpasses that of radionuclides, at around 1.5 Myr. At the same time, Panel (b) shows the evolution of the global rock/ice mass ratio. A large fraction of the water is absorbed/embedded within the rocks. As a result, our initial rock/ice mass ratio of 4, grows to nearly 14. Very little water remains as pure ice. As indicated by Figure \ref{fig:RoWfinal_20-5-0.1-1-1}, most of it is concentrated near the surface (vapour transport) and at around 1.5 km below the surface (liquid transport).
	
\begin{figure}
		{\includegraphics[scale=0.63]{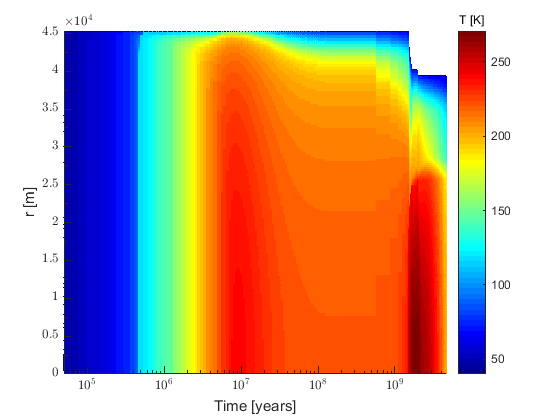}}
	
		\caption{Evolving temperature, depicted in terms of the radial distance from the centre vs. time, for a 50 km ($\sim$45 km, after initial pebble packing compression) radius comet, with the following parameters: formation time is 6 Myr after the formation of CAI, pebble size is 1 cm, mineral fraction is 0.5 and permeability parameter is $b=7$.}
	
		\label{fig:T_50-6-1-0.5-7}
	\end{figure}

	In summary, we have obtained a dense comet nucleus with a global bulk density of about 1730 kg $\times$ m$^{-3}$. It has an intricate structure consisting of a low density, hierarchical pebble surface layer, overlying a dense and homogeneous inner core, consisting of micron-sized particles. The outer layer is composed of anhydrous rocky minerals, whereas the core consists of hydrous minerals. Both layers have a similar water structure in the sense that they are enriched in water ice at the top and depleted of ice at the base. The topmost surface layer might also harbour super- and hyper-volatiles that have accumulated there early-on in the evolution, having been migrated from the inner parts of the comet. The orbit must be Oort-like for the majority of hyper-volatiles to be stable on the time scale of the Solar system. For completion, we also show in Figure \ref{fig:T_50-6-1-0.5-7} the temperature evolution of a 50 km radius comet. Although the general outcome is similar, the notable difference is the occasional presence of a second, Gyr peak in temperature, due to the now important contribution of LLRs, and the retention of high, present-day core temperature. We note however that in this example there is no early aqueous collapse, due to the relatively large formation time and small mineral fraction in this simulation. With early aqueous collapse, the resulting thermal conductivity would have been significantly higher (via greatly reduced porosity), and thus a significant second LLR peak would not have occurred. Overall, we find that a second LLR temperature peak is uncommon, but possible.
    
    \section{Discussion}\label{S:Discussion}   
    
    \subsection{Constraints on Solar system formation models}
    The results described above have important implications for understanding the formation of comets, as well as similar-sized planetesimals (such as Kuiper belt or Ooort cloud objects) in general. As is now widely accepted, planetesimals can be formed by the gentle collapse of a pebble cloud concentrated by the streaming instability \citep{Youdin.2005,Johansen.2007}. This formation process was investigated in several studies and was found to be robust \citep{Johansen.2007,Johansen.2009,Johansen.2011,Johansen.2012,Kato.2012,Rucska.2020,Carrera.2021b}. However, the mean size of the simulated planetesimals lies between $50$ and $1,000 \,\mathrm{km}$ \citep{Simon.2016,Simon.2017,Schafer.2017,Abod.2019}, with the lower end being the result of the finite numerical resolution. Recent studies indicate that smaller, comet-sized planetesimals may indeed form in the solar nebula, but require an enhanced pebble-to-gas ratio \citep{Klahr.2020b,KlahrSchreiber2021}. This in turn implies that part of the gas in the nebula has already dissipated, and therefore that comets with radii below 50 km must have formed relatively late.
    
    It may also be argued that small comets are collisional products of large comets. In a forthcoming study, we will investigate comets with radii exceeding 50 km, which were ignored in this work. However, here we show that even for comets with radii of just 10-20 km, dramatic changes in the interior structure occur. For a large portion of our parameter space, melting of water ice takes place. Under our assumption of aqueous collapse, these comets shrink considerably and their final densities are in excess of 1500 kg $\times$ m$^{-3}$, similar to many asteroids. Thus far, our knowledge of small comets supports typical densities around 500 kg $\times$ m$^{-3}$. It seems unlikely that the latter are collisional products of the former, since catastrophic collisions and gravitational re-accretion of the ensuing fragments would require those fragments to be strictly sub-meter, or else they would have been resolvable by the Consert measurements of \cite{Herique.2019}, and heterogeneity would have been detected in the nucleus of comet 67P/C-G. Further findings of Consert are discussed in Section \ref{SS:Permittivity}. We may conclude, based on 67P/C-G, that small comets are not collisional products of larger comets, but such conclusions should also be taken with a grain of salt, as 67P/C-G is currently the only comet ever studied with such precision.
    
    If indeed small comets are not collisional products, and if splitting is not universal among all comets, during their lifetimes, their size may be determined during the initial gravitational collapse of a pebble cloud. As suggested by \cite{Klahr.2020b,KlahrSchreiber2021}, this process could be time-dependent. Large planetesimals form first, whereas small ones form later (when the gas has started to dissipate) and farther from the Sun. If true, our study could have interesting implications, because it results in evolutionary outcomes that largely diverge from past work.
    
    Recent work on the internal heating of planetesimals focuses on widely varying comet sizes and formation times. E.g., \citet{GolabekJutzi-2021} found that for peak temperatures in excess of 40 K, planetesimals must be 20-40 km in radius, depending on their exact formation time. For temperatures in excess of 80K and 140K, the planetesimals  must be even larger, and also form earlier (as early as 2.5 Myr - see their figure 9 - dashed black curves).
    
    \cite{LichtenbergEtAl-2021} also studied the evolution of the interior of planetesimals. As they consider only formation times $\leq$3 Myrs, their results are not directly comparable to ours, since they use a mineral fraction of 1 (20 times larger than 0.05 in our study, for the same formation times). Under those conditions, they derive high mean temperatures reaching 150 K even for small bodies with a radius of 1 km.
    
    \cite{MousisEtAl-2017} investigated the influence of radiogenic heating on bodies of comparable sizes to this work, but as in the previous two studies mentioned, they did not account for the complex details of pebble structure. For a body with a 1.3 km radius, they found that the presence of amorphous ice implies a formation time of at least 2.5 Myr after CAI, while for a 35 km radius, the formation time increases to 5.5 Myr.
    
    Comparing the aforementioned results to our study, as well as earlier work discussed in Section \ref{S:Intro}, it becomes obvious that the internal structure of planetesimals has a decisive role in determining their evolutionary fate. Our study is the first to consider the implications of pebble structure on the long-term thermo-physical evolution of comets, using highly detailed, empirically derived constituent relations for pebbles.
    
\begin{figure}
		\subfigure[Early formation; mineral fraction=0.05]{\includegraphics[scale=0.62]{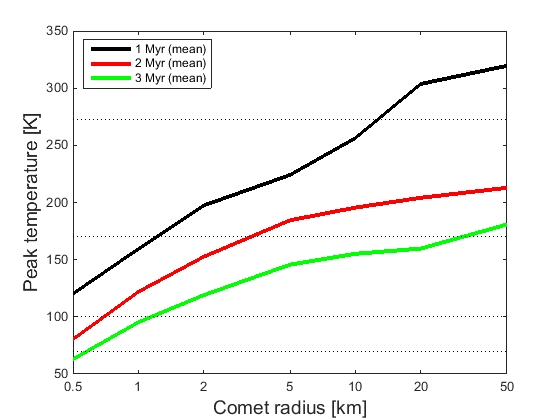}\label{fig:PeakTemperature_early}}
		
		\subfigure[Late formation; mineral fraction=0.5,1]{\includegraphics[scale=0.62]{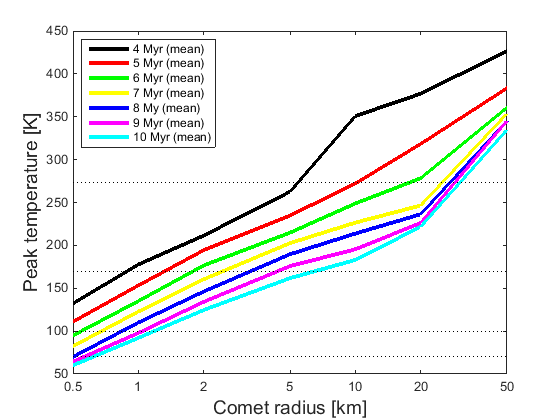}\label{fig:PeakTemperature_late}}
		
		\caption{The maximal internal temperature inside the planetesimal as a function of its radius, given different formation times  (see legend for the colour scheme). Each data point consists of the mean of all the models with different pebble sizes, permeability parameters and (when applicable) mineral fractions. Thus, the solid lines depict the averaged trend. The dotted horizontal lines correspond to the categorical temperatures in Figures \ref{fig:0.1-0.5-1}-\ref{fig:0.05}.}
		
		\label{fig:PeakTemperature}
	\end{figure}

    Figure \ref{fig:PeakTemperature} shows that any comet in our study with a radius exceeding 2 km, reaches peak temperatures in excess of 100 K, \emph{regardless of what model parameters were used}. This implies that contemporary comets must have formed extremely small, or extremely late, or with a tiny mineral fraction, if they were to avoid any differentiation and preserve their pristine composition (noting that this statement ignores the possibility of subsequent collisions). Alternatively, since there is no conclusive evidence that comets are indeed undifferentiated (see further discussion in the next section), it is entirely plausible that they are in fact differentiated in some manner.
    
\begin{figure}
		\subfigure[Early formation; mineral fraction=0.05]{\includegraphics[scale=0.62]{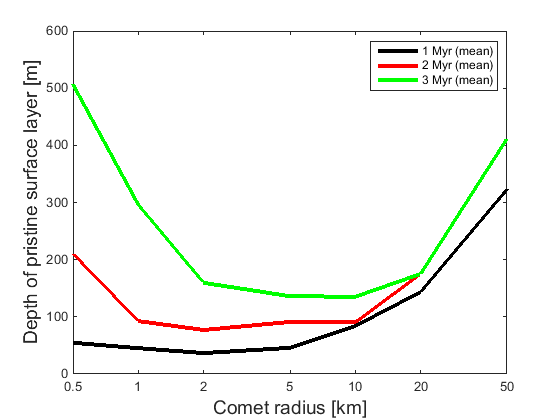}\label{fig:PristineLayer_early}}
		
		\subfigure[Late formation; mineral fraction=0.5,1]{\includegraphics[scale=0.62]{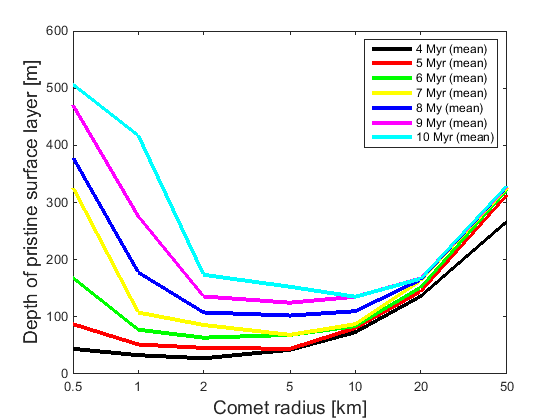}\label{fig:PristineLayer_late}}
		
		\caption{The depth of the pristine ($T<70$ K) surface layer of the planetesimal as a function of its radius for different formation times. The solid lines and their corresponding colours have the same meaning as in Figure \ref{fig:PeakTemperature}.}
		
		\label{fig:PristineLayer}
	\end{figure}

    The latter point means, at the very least, that hyper- and super-volatiles could have been redistributed from the central parts of the comet to its outer portions, where temperatures are sufficiently low to allow for re-condensation. Figure \ref{fig:PristineLayer} shows the depth of the pristine surface layer (temperature never exceeded 70 K) versus the comet radius. We expect the hyper-volatiles to re-condense somewhere within this layer, or else sublimate from the surface early-on (depending on the orbit), and the super-volatiles to condense at its base. For most of our simulated bodies, the pristine layer reaches a depth of 200 m or less, except for the smallest or largest comets. We note that future studies would require superior resolution in order to obtain finer zoning of the numerical grid close to the surface and hence improve the accuracy of our current results.
    
    For comet 67P/C–G, a mean mass loss of at least 1 m per orbit \citep{Patzold.2019,Nesvorny.2018b} and possibly around $\sim$10 m per orbit \citep{Fulle-2021} was observed. Thus, between a few to a few hundred close perihelion passages might be required before the erosion by activity reaches deep enough layers to unveil the super-volatile-enriched zones. Further erosion might then expose a declining gradient of super-volatiles, depending on the characteristic burial depths of each species, until eventually all super-volatiles might be depleted. Such a transition of activity might not take place on human time scales, but this aspect should be considered in more detail in future research. 
    
    One potential example that such a transition is already taking place is for comet 2P/Encke, which is possibly the most thermally evolved comet known. Although 2P/Encke may have experienced a rich dynamical past through its interactions with the giant and terrestrial planets, including long periods of dormancy \citep{LevisonEtAl-2006}, it is today an active comet with a small perihelion distance of merely 0.33 AU and an orbital period of 3.3 yr. At its current (or likely pre-erosion) size, 2P/Encke retains a considerable volume fraction of amorphous ice and other high volatility ices for a large parameter space in Figures \ref{fig:0.1-0.5-1}-\ref{fig:0.05}, in compatibility with its observed high volatility species. Comparisons made for its observed hyper- and super-volatile content between its 2003 and 2017 apparitions yielded very different abundances. E.g, see figure 5 of \cite{RothEtAl-2018} wherein the hyper-volatile species CO, CH$_4$, C$_2$H$_6$ and C$_2$H$_2$ became significantly depleted, while some super-volatile species display the opposite trend. This could be interpreted as a more advanced state of erosion in which less hyper-volatiles and more super-volatiles were released, although we note that the observations were made at different heliocentric distances of $\sim$1.2 AU and $\sim$0.5 AU, respectively, which might also account for these differences. Further investigation is required.
    
    Figures \ref{fig:PeakTemperature} and \ref{fig:PristineLayer} both demonstrate that there is a correlation between the formation time of small planetesimals, the lowering of their peak temperatures and thickening of their pristine surface layers. This in turn can place testable constraints for the hypothesis of \cite{Klahr.2020b,KlahrSchreiber2021}. However, the major uncertain parameter is how large the mineral fraction could be. This single parameter, which according to the Stardust samples is virtually non-existent \citep{Levasseur-RegourdEtAl-2018}, may serve to differentiate between early and late formation histories.
    
    \subsection{\label{sect:pristinity}Are there actually any pristine comets?}
    From the discussion in the preceding subsection, we have already seen that most of the planetesimals simulated in our study undergo essential differentiation. Thus, there are almost no pristine comets expected to exist today if pristinity refers to their internal structure and composition at formation. Figure \ref{fig:PeakTemperature} reveals that (for the parameter space explored by figures \ref{fig:0.1-0.5-1}-\ref{fig:0.05}) a radius of 1 km is sufficient to drive the peak temperature above 70 K, regardless of the mineral fraction, at which point both the hyper- and super-volatiles would sublimate, diffuse outwards and either re-condense closer to the surface or escape into space. The details of this process are beyond the scope of this paper and will be considered in a follow-up work. For planetesimals with radii $\ge$2 km, regardless of their formation time or mineral fraction, the peak temperatures exceed $100$ K so that the super-volatiles (primarily CO$_2$) diffuse outward and re-condense below the pristine surface layer (see Figure \ref{fig:PristineLayer}). The above does not apply to comets that both have an extremely small mineral fraction and form much later than 3 Myr (i.e beyond the range of Figure \ref{fig:0.05}). These can have limit radii larger than 1 and 2 km, respectively.
    
\begin{figure}
		\subfigure[Early formation; mineral fraction=0.05]{\includegraphics[scale=0.62]{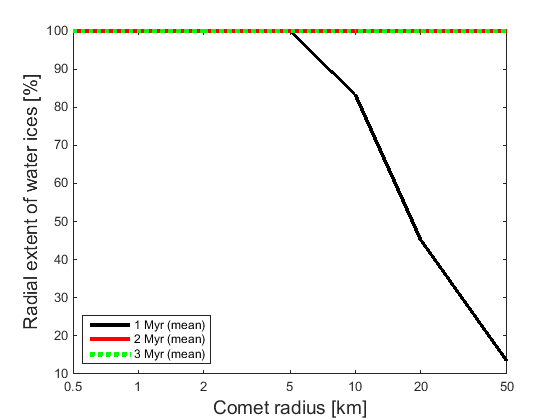}\label{fig:CrystallineLayer_early}}
		
		\subfigure[Late formation; mineral fraction=0.5,1]{\includegraphics[scale=0.62]{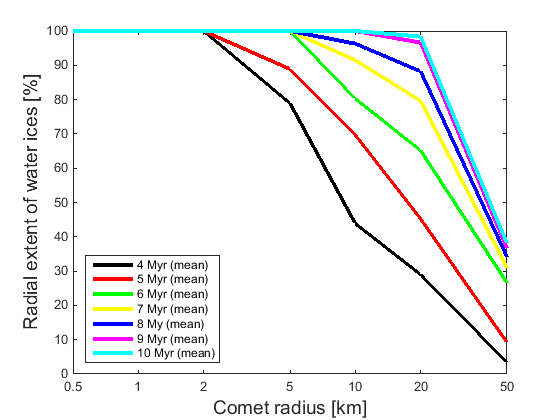}\label{fig:CrystallineLayer_late}}
		
		\caption{The radial delineation of either amorphous or crystalline water is shown, as a function of comet radius for different formation times. The solid lines and their corresponding colours have the same meaning as in Figure \ref{fig:PeakTemperature}. They depict the radial fraction that contains crystalline water ice, starting from the surface (e.g 100\% covers the full radial extent, from surface to centre).}
		
		\label{fig:CrystallineLayer}
	\end{figure}

    For the case of water ice, we explicitly followed its re-distribution. E.g., in Figure \ref{fig:final10-5-1-1-7}, when the temperature exceeded a threshold value, water ice began sublimating, migrated outward and re-condensed in a narrow radial zone in which its local density was enhanced by a factor of 3-4 while in turn the local porosity decreased to less than 50\%. In contrast, the water-ice interior density decreased by roughly a factor 2. In Figure \ref{fig:CrystallineLayer}, we show the radial delineation of the various phases of water ice (i.e either amorphous or crystalline) for the entire suite of simulations. A value of 100\% means that water ice extends all the way from the surface to the centre of the comet, whereas a value of 50\% means that it extends from the surface to the mid point between the surface and the centre (i.e. half of the comet, radially, is devoid of water). The results show a correlation between the comet's size and redistribution of water, but always only extremely large comets are able to fully deplete water from some portion of the inner core. This last sentence is however incorrect for comets that have a mineral fraction of 0.05. Their water ice distribution is uniform, unless their formation time is around 1 Myr. But note that the delineation of crystalline as opposed to amorphous ice is still radially distinct, almost in all cases.
    
    For the super-volatiles, we expect qualitatively a similar behaviour. Due to the differentiation, the radial abundances of the volatiles (relative to each other and relative to the refractories) vary considerably. In this sense, there are no pristine comets, because all radial zones of today's planetesimals are either enriched or depleted by hyper-volatiles, super-volatiles or volatiles. Thus, measuring the abundance ratios in comets might not reflect their original values.
    
    To contrast, the current view of comets is that they are actually undifferentiated. This view is motivated by various observations of comet splitting events, and in particular the splitting of Shoemaker-Levy-9, which is actually the only case of splitting where the root cause is completely undisputed (tidal splitting) and the process is fully understood from theory \citep{Boehnhardt-2002}. At a distance of 5.4 AU, Shoemaker-Levy-9 tidal fragments remained active, which sparked the (still prevalent) general notion that the progenitor comet must have been both pristine (because crystalline ice could not generate the observed dust activity) and homogeneous. However, in fact, the progenitor does not have to be homogeneous. Tidal disruption models indicate that it can be radially differentiated into several compositional layers. It is enough that only the outermost layers contain materials capable of triggering the dust activity. This comes from the work of \cite{HahnRettig-1998}, and was also demonstrated by \cite{MalamudPerets-2020a,MalamudPerets-2020b} via more modern and advanced simulation tools. Such studies demonstrate that the progenitor comet undergoes tidal elongation and subsequent collapse of a gravitationally self-confined tidal stream. The ensuing tidal fragments contain material which spans several of the progenitor's radial zones. They can be fully or partially assembled from the progenitor's outer layer materials, so regardless if the Shoemaker-Levy-9 progenitor was radially heterogeneous, there should have been material capable of trigerring the observed dust activity in the tidal fragments. Thus, based on Shoemaker-Levy-9, the most well-understood split comet, there is actually no conclusive evidence that most comets are globally homogeneous.
    
    Our current study rather implies that future exploration of comets, both in-situ and remote, could benefit from focusing on small comets. Figures \ref{fig:PeakTemperature} and \ref{fig:PristineLayer} show that the smallest comets in our sample, with radii around 0.5 km, are also the most pristine. With the exception of hypervolatiles, they might be the most suitable for accurately representing primordial composition. However, note that small comets may in principle also be collisional products or fragments ensuing from splitting of a much larger comets. Hence, it has to be substantiated, if possible, that their current size actually reflects on their original size.
    
    \subsection{Parameter sensitivity}
    We first consider the results of Section \ref{SS:4-10Myr}, where the mineral fraction is either 50\% or 100\%. Figures \ref{fig:PeakTemperature} and \ref{fig:PristineLayer} show the mean evolutionary outcomes, based on different model realisations, with varying pebble sizes, mineral fractions and permeability coefficients. In order to evaluate the relative importance of each of these three parameters, we must examine Figures \ref{fig:0.1-0.5-1}-\ref{fig:1-1-7} more closely.
    
    First, consider Figures \ref{fig:0.1-0.5-1}-\ref{fig:0.1-1-7}, which show the evolutionary outcomes for the smallest pebbles, and Figures \ref{fig:1-0.5-1}-\ref{fig:1-1-7} for the largest pebbles. While in the former group there are aqueously altered comets in all 4 figures, in the latter group they are only found in 2 out of 4 figures. Clearly, for large comets, large pebbles induce lower peak temperatures than small pebbles, i.e reaching the water melting temperature more easily. To contrast, small comets are actually more evolved when the pebbles are small. This seemingly opposite result is understood from Equations \ref{EQ:hertz_factor} and \ref{EQ:RadiationTerm}. The radiative term of the effective thermal conductivity scales with the pebble size $r_{\rm peb}$, and the solid-state term goes as $r_{\rm peb}^{-1/3}$. Hence, for large comets, where temperatures are high enough during much of the evolution, the radiation term dominates and conductivity strongly correlates with the pebble size. For small comets, the opposite is true, and conductivity anti-correlates with the pebble size. While the pebble size also has other indirect effects, the fact that heat is transported more (less) efficiently with larger pebbles, naturally explains why it is harder (easier) for the temperatures to build up in large (small) comets.
    
    Next, consider the mineral fraction. It is trivial to think that by doubling the radiogenic materials, we increase the available internal heat, and indeed as we look at each pair of figures with identical parameters except mineral fraction, we recognise this trend very clearly. E.g, when one compares Figure \ref{fig:0.1-0.5-1} to Figure \ref{fig:0.1-1-1}, the formation time that results in either aqueously altered large comets, or fully pristine small comets, increases from 5 Myr to 8 Myr.
    
    Of the three free parameters, the permeability coefficient seems to be the least important for the evolutionary outcomes. In principle, in a more permeable medium mass is transported out more easily, which in turn implies that heat transport by advection is more important. A permeable comet thus heats less than its low permeability counterpart. Each successive figure pair among Figures \ref{fig:0.1-0.5-1}-\ref{fig:1-1-7} shows this trend plainly. However, since our model only considers the physical transport of water and not super- or hyper-volatiles, one indeed notices that such differences are manifested only in relatively large comets in which water transport actually takes place. Overall, the fact that this coefficient has a more limited impact on our results is fortunate, since as explained in Section \ref{SS:FreeParameters}, its exact value is currently an open question.
    
    To contrast, the results of Section \ref{SS:1-3Myr} investigate the early formation between 1-3 Myr. If the mineral fraction is as low as 5\%, the models are not very sensitive to either pebble size or permeability. There is however a strong dependence on the formation time and comet size. Similarly to Section \ref{SS:4-10Myr}, only extremely large comets whose radii are larger than about 20 km are capable of attaining water melting temperatures. However the formation must be exceedingly early, at around 1 Myr after the formation of CAI. 
    
    We note that it is also possible that the actual mineral fraction is in between 5\% and 50\%, or that different comets have different mineral fractions, which opens up a large array of possibilities. These results underscore the significance of this parameter, which remains to this day highly unexplored and unconstrained, regardless of the results of the Stardust mission. Our study thus strongly advocate the necessity for repeating in situ missions for the return and analysis of cometary samples.
    
    \subsection{Nucleus spin-down by vapour transport}\label{SS:Spin-down}    
    Consider the initial, pre-differentiation angular momentum of a homogeneous comet nucleolus $I \omega$, where $I$ is the moment of inertia and $\omega$ the angular speed. It follows from conservation of angular momentum that the post-differentiation angular momentum $I' \omega'$ leads to a change in spin, such that $\omega'=(I/I') \cdot \omega$.
    
    We have shown in Section \ref{S:ParameterStudy}, and demonstrated in Section \ref{SS:WaterVapor} that under certain conditions comets may undergo water differentiation through vapour transport. This means that the inner part of the comet loses water, hence its density decreases, while the outer part gains additional water, hence its density increases. Unlike hyper- and super-volatiles, water is a significant non-refractory component in comets (see Section \ref{SS:Pristine}), therefore vapour-driven differentiation has the potential to change the moment of inertia considerably. From an initial value of 0.4, for an assumed homogeneous sphere, the new structure (lighter core underlying denser mantle) necessarily results in a moment of inertia factor greater than 0.4. In turn the angular speed from the above equation must decrease - the nucleus spins slower than before. 
    
    For example, we calculate the moment of inertia factor of the comet in Figure \ref{fig:10-5-1-1-7} (Section \ref{SS:WaterVapor}) to increase by merely $\sim$1\% as a result of water redistribution. In this case spin-down is thus relatively unimportant. In principle, for comets with a larger initial water content, vapour transport could lead to a more significant increase in moment of inertia factor and therefore also greater spin-down.
    
    We note that in active comets, changes in the nucleus spin can occur on much shorter time scales simply by sublimation-induced torques, driven by insolation, as indicated e.g. by \cite{KellerEtAl-2015} and references therein. For active comets, the spin state is therefore unlikely to be related to nucleus spin-down by vapour transport. In other words, we expect the comet to have memory of such primordial processes, perhaps, only if it has remained inactive and far from its host-star since its epoch of internal vapour migration. In the next section we discuss the implications of another potential spin-changing effect triggered by aqueous alterations, which significantly reduce the comet size due to pebble collapse, and therefore can have the opposite effect of spinning the comet up.
    
    \subsection{Aqueous collapse, spin-up and possible instability}\label{SS:Spin-up}
    We have shown in Section \ref{S:ParameterStudy} that large comets are capable of reaching temperatures that lead to water melting, and in turn drive global scale aqueous alterations via serpentinization reactions that sweep through the body. This outcome is robust and extends to a large portion of our investigated parameter space. These processes were further discussed in Section \ref{SS:AqueousAlteration}, and were shown to significantly decrease the initial radius of the comet, under the assumption of aqueous collapse of the pebbles. We are currently planning laboratory experiments to validate the conditions of aqueous collapse in a forthcoming study. Assuming aqueous collapse, we find the radial decrease in our simulation suite to be in the range 16-32\%. However, in all but a few marginal cases, the decrease is characteristically around 30\%. As a rule-of-thumb we take this to be the typical size shrinkage of an aqueously altered, pebble structure comet.
    
    We investigate two related implications of radial collapse. First we rule out the possibility that contribution from gravitational potential energy is important. In principle, since it is inversely propositional to the radius, gravitational potential energy could be released as a result of radial collapse. Our code however includes a precise calculation of this effect (see section 2.6 of \cite{MalamudPrialnik-2013}), and we can confidently say that the energy gained in this way is entirely negligible compared to other energy contributions, and may be ignored.
    
    The second implication of radial collapse is spinning up the comet's rotation. Following the analysis in Section \ref{SS:Spin-down}, the angular speed changes since the momentum of inertia of the aqueously altered comet, $I'$, greatly decreases. The decrease is two-fold. First, as shown in Section \ref{SS:AqueousAlteration}, internal processes lead to a comet that has a compacted core underlying a much lighter pebble mantle. In such a configuration the moment of inertia factor is lower than 0.4. Far more important however is the change to the moment of inertia due to radial collapse. To order of magnitude, the moment of inertia of a simple sphere goes as $M \times R^2$, where $M$ is its mass and $R$ its radius. While the mass remains constant, the radius $R$ decreases. Taking 30\% to be the characteristic radial decrease, the new moment of inertia is roughly halved. This means that as a rule-of-thumb, the nucleolus spins twice as fast as it did before the radial collapse.
    
    If the initial rotation of the comet is already considerable, spin-up might have interesting consequences. The alleged spin-up takes place during a phase in the comet's evolution in which it experiences sudden and relatively rapid radial decrease due to motion of liquid water and pebble collapse in its internal portions. Hence, one might speculate that a spinning, contracting body, with a mobile interior would eventually settle into a shape that is far less irregular than it was prior to the collapse, as follows. 
    
    In the limiting case of a MacLaurin Spheroid, faster rotation might yield an oblate, fully spheroidal shape. However, our model predicts the appearance of liquid water partially filling only the internal refractory (and probably still rigid) matrix, underlying an addition cold, outer pebble mantle. We might speculate that this combination would reduce shape irregularity but would not eliminate it completely. 
    
    An even more interesting possibility is that spin-up could lead to rotational breakup and splitting of the nucleus. The limit rotation period as a function of nucleus size is given by equations 19, 31 or 33 in \cite{Davidsson-2001}, depending on the assumed nucleus shape. The parameters required are the comet density and tensile strength, and in the case of an oblate shape also the flattening parameter. \cite{KokotanekovaEtAl-2017} for example, plot the observed rotations of multiple comets as a function of their size (see their figure 61), and overplot the theoretical \cite{Davidsson-2001} curve for a choice of parameters that agrees with comet 67P/C–G. It can be seen that for comets with radii in the range 3-10 km, the rotation limit is almost constant around 6 h, however we emphasize that this limit curve is derived under the assumption that the density is as low as that of 67P/C–G. Given our results, comets in this size range rarely become aqueously altered. Hence, they do not collapse, and their densities never deviate much from their initial values. It is therefore not surprising that no comet lies below this theoretical curve.
    
    Larger comets are however more susceptible to aqueous collapse. Their post-collapse densities are a few times larger compared to their pre-collapse values, and therefore the breakup limit curve needs to be re-derived based on a judicious density. Using the typical density from Section \ref{SS:AqueousAlteration} and equation 33 from \cite{Davidsson-2001}, we find that the breakup rotation period is much smaller, around $\sim$2.5 h - very similar to the known asteroid breakup limit. We showed above that the rotation period should be approximately halved by a characteristic aqueous radial collapse. It therefore seems that any primordial spin under $\sim$5 h might later result in spin-up, triggering a rotational breakup and splitting of the nucleus (i.e. by driving the comet below the breakup limit curve). Since empirical evidence however places the average comet spin at around 15 h \citep{Whipple-1982}, it seems that this scenario is possible, yet improbable.
    
    Despite the conclusion of the previous paragraph, there might still be some frail evidence in support of spin-up and splitting. It was previously argued that one of the largest comets known, Hale-Bopp, has a satellite \citep{Sekanina-1997}. Hale-Bopp has a radius above 30 km, hence its original radius would have been approximately in the range 40-50 km and certainly large enough to result in aqueous alterations and trigger pebble collapse across a huge range of model parameters. Additionally, the Kuiper belt possesses many objects that are large enough to be aqueously altered. It is also known to harbour a considerable fraction of binaries \citep{NollEtAl-2020}, hence these facts might be linked. We note however that there are certainly other explanations invoked for this binarity besides splitting \citep{GoldreichEtAl-2002,FunatoEtAl-2004,AstakhovEtAl-2005,SchlichtingSari-2008,PeretsNaoz-2009,NesvornyEtAl-2010}. Furthermore, for radii exceeding 50 km, compaction by self-gravity might predate and therefore annul aqueous collapse, potentially altering our simulation outcomes. We are investigating these aspects in more detail in a forthcoming study.
    
    \subsection{No internally driven outbursts or implosions}
    Build-up of pressure as a result of insolation could lead to dust ejection. If the gas pressure exceeds the tensile strength of the near-surface dust particles (either pebble clusters, single pebbles or sub-pebble dust), an ejection is triggered, as was investigated by \cite{FulleEtAl-2020} and \cite{GundlachEtAl-2020}. Here we apply a similar check, however since we only consider inactive comets whose surfaces never heat considerably by insolation, our current investigation focuses rather on the bulk of the comet instead, and not the surface layers. In other words, we examine (temporally as well as spatially) whether the gas pressure \emph{anywhere} inside the comet can ever exceed the pressure of self gravity by the overlying layers. Our simulations show that this never occurs. For aqueously altered comets, the ratio between the gas pressure and gravity appears to be largest, exceeding that of less thermally processed comets. Still, hundreds of meters below the surface, where this ratio is characteristically found to be maximal, it does not amount to more than a few \% - and thus never exceeds the threshold that leads to outburst or implosion of the comet.
    
    \subsection{Observed dust-to-volatile ratio}
    The subject of dust-to-volatile ratio in comets has been recently reviewed by \cite{FulleEtAl-2019} and \cite{ChoukrounEtAl-2020}. These reviews underscore the fact that we cannot depend on such measurements in order to constrain the origin or true nature of the comet's composition, as this ratio changes during different points in time: given various modes of activity, temporally variable illumination conditions, the realisation that over 80\% of the refractories fall back on the surface whereas the lost material is depleted by the same amount -- hence not reflecting the true nature of the nucleus, the recently considered effect of compositional in-homogeneity amongst the pebbles themselves \citep{Fulle-2021} and finally by the fact that the nucleus itself may be heterogeneous.
    
    Our study also strongly advocates in favour of the last point. We have demonstrated that radioactive heating inside comets leads to considerable changes and diversification of their internal composition. Due to the evolving temperature gradients in the interior of these bodies, the inner volatiles sublimate and diffuse into the outer colder layers, where recondensation occurs (see e.g. Panel (a) of Figure \ref{fig:final10-5-1-1-7}). Only hyper-volatiles may escape recondensation, depending on the comet's orbit. This effect leads to the formation of ice-enriched shells around volatile-depleted or at least volatile-reduced cores. The burial depth of each volatile species follows a complex radial gradient, depending on the precise attributes of the nucleus such as its size, formation time and mineral fraction. Without precise knowledge of these and other uncertain parameters, one could not calculate such gradients within absolute certainty, however two general statements indeed apply. The first is that the dust-to-volatile ratio of the outer active layers of a comet, does not reflect on the original bulk ratio at the time of the comet's inception. It is both lower (since volatiles are enriched), and also includes release of various impurities, as was discussed in Section \ref{SS:SuperVolatiles}. The second is that it is temporally variable, and given not only to changes on an orbital time scale, but perhaps mostly on the time scale of many orbits, as compositional gradients are incrementally exposed by erosion.
    
    This means that most comets that we can investigate by spacecraft missions will possess a dust-to-volatile gradient with higher ratios in the interior, and local minima towards the outer layers. Since most comets are small, this refers primarily to hyper- and super-volatiles (not water), and does not depend on the mineral fraction (see Figure \ref{fig:PeakTemperature}). Spacecraft missions capable of sensing the dust-to-ice ratio of comets will also conclude various values of the dust-to-volatile ratio depending on their sensing range. As an example, the complicated derivation of the nucleus dust-to-volatile ratio from the dust-to-gas ratio only provides a measurement of the surface layer mixing ratio, whereas radio- and microwave-science instruments are capable of sensing into deeper layers (e.g., Consert on-board the Rosetta/Philae spacecraft). Depending on their wavelength, these instruments will be able to provide data on the dust-to-volatile ratio of the interior. However in all cases, proper thermal modeling support is required in order to interpret how the local distributions (that is, the integrated mass losses over an orbital time span) translate into a global dust-to-volatile ratio.
    
    It should be noted that most sub-kilometre objects that have also formed relatively late (i.e formation time being in the mid-upper range of Figures \ref{fig:0.1-0.5-1}-\ref{fig:0.05}, respectively) will not be affected much by their thermal evolution. For these objects, spacecraft measurements can be used to infer a global value of the dust-to-volatile ratio with much greater certainty than for large comets (however see the caveats in Section \ref{SS:Pristine} about comet collisions and splitting).

    \subsection{Inert asteroids on typical cometary orbits: past, volatile-depleted cometary cores?}\label{SS:inert}
    
     123 objects listed in the infrared asteroid catalogues (I–A–W) have been identified as so-called asteroids in cometary-like orbits (ACOs) \citep{Kimetal2014,Usuietal2014}. These objects possess orbits similar to Jupiter-family comets (JFCs), or Halley-type comets (HTCs), but they do not show any sublimation-driven activity phenomena. Two different classes of ACOs, high-albedo objects with a geometric albedo of $p_v > 0.1$ and low-albedo ACOs with a mean geometric albedo of $p_v = 0.049\pm0.020$, have been detected. Because the low-albedo ACOs hava a very similar mean geometric albedo compared to JFCs (for comparison, the geometric albedo of JFCs is $p_v = 0.046\pm0.020$), this second class is also renown as potentially dormant comets (PDCs).
    
    For a better comparison of the PDCs to the different comet families (JFCs and HTCs), the PDCs are typically divided into two subgroups, one group having Tisserand parameters in the range $2<T_{\rm J}<3$ (PDCs in JFC-like orbits) and another group having Tisserand parameters satisfying $T_{\rm J}<2$ (PDCs in HTC-like orbits). A comparison of the PDCs in HTC-orbits with their active counterparts shows that PDCs have a shallower size distribution \citep[see Fig. 3 in][]{Kimetal2014}. In particular for radii between $2.5$ and $50 \, \mathrm{km}$, the relative amount of inactive objects is increased, which indicates that objects in this size range deplete their surface volatile reservoirs faster without creating new active areas. The same effect can be observed when comparing the PDCs in JFC-like orbits with JFCs. For radii larger than $2\, \mathrm{km}$, an increased amount of inactive objects can be observed, whereas smaller objects tend to be active more often, or are able to sustain their activity.
    
    Thus, the $2\, \mathrm{km}$-radius border seems to indicate the critical radius at which comets tend to diminish their ability to create observable cometary activity. Earlier work by \cite{TancrediEtAl-2006} arrived at a similar conclusion (invoking a truncation radius of around 3 km), and offering a possible solution in which large nuclei are able to form an insulating dust mantle faster than small nuclei. If gravity is indeed the root cause, one might expect a gradual (rather than a sharp) transition going from small to large objects, which seems to be the case according to their figure 16. New studies however that it is difficult to form an insulating dust mantle in an active comet nucleus \citep{FulleEtAl-2020,CiarnielloEtAl-2022}.
    
    Here we therefore consider two other possible explanations for this turn-off point in the context of our model. The first possibility is exemplified in Figure \ref{fig:CrystallineLayer}, in which we show that large comets (radii $>\sim$ 10 km) can be heavily modified by thermal alteration, which causes a substantial amount of the water to be relocated to a distinct shell close to the surface. If this water-ice reservoir is sufficiently depleted, cometary activity could expire, which potentially explains the existence of PDCs in this size range. I.e., if PDCs experienced complete water depletion in their interior cores (or else just severe water depletion, as in Panel (a) of Figure \ref{fig:final10-5-1-1-7}, for a particular set of model parameters) and if their exterior water-ice shells are already exhausted, they would not be able to restart prolonged activity (however short term activity could be achieved in non-depleted cores by coming closer to the Sun or by removing surface layers). Then, the whole object contains very little water ice with an insulating dust mantle, or even no water ice at all. In this case, the classification as potentially dormant comets would be misleading. These objects are permanently extinct (not dormant) comets.
    
    The second possibility, since water depletion or even significant water reduction in the interior requires a radius considerably larger than the aforementioned 2 km turn-off radius, is that super-volatiles are the root cause behind this observed trend. In many comets, the orbit (perihelion distance) might be such that strong activity requires the presence of super-volatiles. Our results indicate that for comets with radii in excess of 2 km, a considerable volume fraction is warmed past the 70-100 K threshold that allows the release of impurities. This means that for these comets the interior will be significantly devoid of super-volatile impurities, which would have been released already during the early processing of amorphous water or CO$_2$ ices, and subsequently transported close to the surface. After activity erodes the super-volatile-rich zones which were closer to the surface, a comet might not be able to resume strong activity in their absence. This might lead to the dichotomy between small and large comets, if the former never lost their bulk impurities in substantial amounts, and thus they can remain active irrespective of how much of the nucleus erodes.
    
    For completion, however, we also note that comet splitting offers an independent explanation for increased activity by smaller comets. Comets are known to split into several active fragments during their journey through the Solar System \citep{Boehnhardt-2002}. Splitting can expose fresh areas, which would enhance the overall activity. After some time, however, the volatile-rich surface of the ensuing fragments might again be replaced by a dust-rich surface, and the activity pattern might return to similar pre-splitting levels. It is therefore not clear if splitting alone can explain the observed trend, and in particular why the turn-off radius is 2 km and not some other value. Although \cite{TancrediEtAl-2006} provide one possibility, we additionally suggest that radial compositional differentiation may be complementary in explaining the observed trend.

    \subsection{Permittivity}
    \label{SS:Permittivity}
    The Consert instrument onboard the Rosetta orbiter and the Philae lander measured a permittivity contrast between shallow layers (0...25 m depth; $\varepsilon=1.70-1.95$) and deeper layers of comet 67P/C-G \citep[25...150 m depth; $\varepsilon=1.20-1.32$; see][]{KofmanEtAl2020}. It was already argued by \citet{KofmanEtAl2020} that there are at least two mechanisms that in principle could explain this permittivity contrast: (i) a decreased porosity in the  sub-surface due to vapour transport induced by insolation; or (ii) composition changes with depth. The latter might be caused by radiogenic-heat-induced mobility of the volatiles as investigated in this paper. However, as described in Section \ref{sect:pristinity}, the re-condensation of volatiles will occur around the pristine layer, depending on their respective condensation temperatures. As we have chosen an exemplary 70 K as a proxy for the depth of the pristine layer (see Figure \ref{fig:PristineLayer}), water ice will condense at greater depths, super-volatiles will condense at the base of this layer and hyper-volatiles might condense closer to the surface, where they might or might not be retained depending on the comet's incipient orbit.
    
    Bearing this in mind, a comparison to Figure \ref{fig:PristineLayer} shows that condensation-enhanced densities as shallow as observed by Consert (0...25 m) are unlikely to be  caused by water ice. However, super-volatiles (e.g., $\mathrm{CO_2}$) might have been (a) accumulated in the observed depths after being relegated from the comet's inner portions, where (b) they are now in turn depleted, and (c) the pristine surface layer also retains its original super-volatiles. Taking into account that comet 67P/C-G has already been eroded by previous perihelion passages at its current orbital elements, if comet 67P/C-G formed at or close to its current size and as early as 4-5 Myrs after CAI (or $\sim$1 Myr for 5\% mineral fraction), the above scenario could be plausible. A quantitative comparison between the  permittivities predicted by our planetesimal-evolution model and the observations of Consert are, however, beyond the scope of this work and also require detailed vapour-transport treatment.
    
    How could the vapour transport induced by radiogenic heating be observed in future missions to primitive bodies of the Solar System? As shown by Consert, radar techniques are capable of penetrating hundreds of meters deep through highly porous granular matter \citep{KofmanEtAl2020}. The permittivity of \emph{solid, non-porous} water ice (at 273 K) has a plateau at $\varepsilon \approx 3$ for frequencies between $\sim 10^5$ Hz and $\sim 10^{12}$ Hz \citep[see figure 1 in][]{Artemov2019}, but rises strongly towards smaller frequencies. The static permittivity of low-temperature water ice is $\varepsilon > 200$ for $T=100$~K \citep[see figure 4 in][]{Artemov2019}. As the bulk permittivity of the refractory porous material of comets is $\varepsilon \approx 1.3$ \citep{KofmanEtAl2020}, high-frequencies radar faces the challenge to detect a rather small permittivity contrast when entering the ice-enriched layer of the body. Ultra-low frequency radar might be a better choice, because of its higher penetration depth and the much higher permittivity contrast between water ice and the refractory components. With such a method, the greater depths of the pristine surface layers of comets could be probed and compared to thermophysical planetesimal-evolution models like the one described in this paper.

    \section{Summary}\label{S:Summary}
    In this paper we bring forth a novel framework for the evolution of comets, under the assumption that they are initially accreted from pebbles. The ultimate effect of pebble structure is to greatly reduce the effective thermal conductivity. We use established empirical relations in order to determine the conductivity, while the comet undergoes thermal and compositional changes as it evolves from its natal state to the present day. Our model also newly considers two effects which were not included in past studies: (a) the compression of pebble inter-porosity by self gravity, for which we performed a dedicated lab experiment, and (b) the full collapse of pebble inter-porosity by aqueous alterations (when applicable), while accounting for the energetic, physical and thermal consequences of the ensuing chemical hydration reactions between liquid water and initially anhydrous rock.
    
    Although our understanding of comet nuclei has grown tremendously over the last few years, primarily through the numerous discoveries made by the Rosetta spacecraft and its array of instruments -- there is still a great deal of uncertainty in the actual relations governing the inner workings of comets. Hence, we explore a large space of model realisations, including various pebble sizes and permeability coefficients. We also raise the possibility that, in similarity to comet 67P/C-G, of all the refractory materials inside a comet, only the minerals may contribute to radiogenic heating, potentially lowering the fraction of radionuclides. Lastly, we consider comets that greatly vary in size and in formation time. These last three parameters are pivotal, and generally determine the course of the evolution.
    
    This parameter space is explored self-consistently via SEMIO, a 1-dimensional thermo-physical code for evolving icy bodies, now newly modified to include the aforementioned pebble-related aspects and empirical relations. We broadly find that when composed of pebble piles, comets may attain much higher temperatures than in any previously published study (when comparing to the same formation time and mineral fraction). 
    
    The primary conclusion of our study is that only extremely small, sub-kilometre comets are truly pristine (in the sense that they may not mobilize anything but hyper-volatiles). This conclusion is robust for any combination of the free parameters in our study, as long as the formation time is not too small. Larger than kilomtre-sized comets may still be pristine, however it requires (in addition to a large formation time) that their mineral fraction is negligible, underscoring its importance. The only available data regarding the isotopic composition of cometary dust exclude that $^{26}$Al was ever present in comet 81P/Wild 2 \citep{Levasseur-RegourdEtAl-2018}, consistent with this requirement. On the other hand, if the rocky materials contribute radionuclides in similar proportions to meteoritic abundances (as \emph{ubiquitously} assumed in all previous thermophysical studies of comets and other icy objects), pristine comets must both be sub-kilometre and form very late ($>9-10$ Myr). This work emphasises the need to better constrain the mineral fraction in comets and other icy bodies across the Solar system. As we investigate a range of comet sizes and formation times, our simulations often result in the processing and mobilization of at least super-volatiles materials, if not water vapour, and eventually even liquid water. The common assumption that comets probe the primordial state of the Solar nebula is therefore often violated.
    
    By this parameter study, we identify four evolutionary archetypes, attempting to categorise pebble-accreted comets according to general classes: (a) pristine comets, which are extremely small and relatively late to form, and which do not experience internal processing of anything but hyper-volatiles; (b) somewhat larger, but still relatively small comets that are capable of at most mobilizing super-volatile ices, or release impurities inside CO$_2$ and amorphous ice; (c) intermediate-sized or relatively large comets which may attain sufficient temperatures in order to mobilize water vapour; and (d) large or extremely large comets that may attain the water melting temperature. These objects undergo major physical, compositional and energetic changes.
    
    There are several potential implications of pebble structure for interpreting the formation of comets in the Solar system. Comets are typically small objects, rarely exceeding radii above 10 km. If pebble comets form through gravitational collapse, they may be either primordially small or else collisional products of larger comets (with radii exceeding $\sim$50 km initially). Our study hints that the first scenario is the most probable, since we show that extremely large comets are typically aqueously altered, and have a dense nucleus, such that their collisional products will not have the observational properties of present-day comets.
    
    We also discuss the early spin-up or spin-down of comets as a result of physical and compositional modifications triggered by radiogenic heating, which change their moment of inertia. Spin-down is deemed to be of typically negligible importance while spin-up seems to be a robust signature of large aqueously altered comets when we assume that their pebble structure collapses, roughly halving their rotation period, which may even result in comet splitting if their initial rotation is already fast.
    
    More implications of pebble structure shed light on the present day composition of comets. We discuss the significance of the internal redistribution of volatiles. The first implication is that the measured dust-to-volatile ratio, even if it is integrated over an entire orbit, does not provide us with the bulk, primordial dust-to-volatile ratio of the comet. This is because volatiles can migrate outwards and concentrate closer to the surface, essentially modifying the initial, radially uniform distribution. However without knowledge of how quickly the comet formed, the exact abundance of radiogenic material and especially how much of the nucleus' external layers have already been eroded by past activity, we cannot attempt to guess the primordial ratio based on the comet's differentiation history as derived from detailed thermo-physical modeling. The second implication is that volatile-depleted cometary cores may eventually lead to cessation of comet activity (i.e. the ongoing ejection of pebbles stops and a lag deposit forms), rendering the comet perpetually extinct.
    
    In conclusion, this study emphasises the importance of pebble structure to the formation and development of comet nuclei. We conclude by suggesting that the delineation of volatiles might be observable by measuring the permittivity of comet nuclei, in similarity to comet 67P/C-G. We also advocate focusing future research efforts to the smallest possible comets, as they might be the only true primordial cometary relics in the Solar system.

    \section{Acknowledgements}\label{S:Acknowledgements}
    We wish to thank Marco Fulle for a thoughtful review of this manuscript, greatly improving its quality. We also thank Noria Brecher for her help in performing the ice-pebble compression experiment. This work was funded by the Niedersächsisches Vorab in the framework of the research cooperation between Israel and Lower Saxony under grant ZN 3630. HBP acknowledges support from the Minerva center for life under extreme planetary conditions. AL is funded by the Deutsche Forschungsgemeinschaft (DFG, German Research Foundation) under grant 278211407 in the framework of the research unit "Debris Disks in Planetary Systems" (FOR2285).
	
	
	
	
	\bibliographystyle{icarus2}
	\bibliography{bibfile}

\end{document}